\documentclass[conference]{IEEEtran}
\makeatletter
\def\ps@headings{%
\def\@oddhead{\mbox{}\scriptsize\rightmark \hfil \thepage}%
\def\@evenhead{\scriptsize\thepage \hfil \leftmark\mbox{}}%
\def\@oddfoot{}%
\def\@evenfoot{}}
\makeatother 

\usepackage{graphicx}
\usepackage{subfigure}
\usepackage{color}
\usepackage{epsfig}
\usepackage{amssymb}
\usepackage{latexsym}
\usepackage{amsmath}

\newcommand {\C} {{\rm I\kern-5.5pt C}}

\newcommand{\bP}[1]{{\mathbb{P}}\left[{#1}\right]}
\newcommand{\bE}[1]{{\mathbb{E}}\left[{#1}\right]}

\newcommand{\1}[1]{{\bf 1}\left[#1\right]}       

\newcommand{\fsquare}{\vrule height6pt width7pt depth1pt}   
\newcommand{\myproof}{{\hfill \\ \bf Proof. \ }}           
\newcommand{\myendpf}{\hfill\fsquare \\[0.1in]}             

\newtheorem{theorem}{Theorem}[section]

\newtheorem{lemma}[theorem]{Lemma}
\newtheorem{proposition}[theorem]{Proposition}
\newtheorem{corollary}[theorem]{Corollary}

\begin{document}

\title{Modeling the pairwise key distribution scheme
       in the presence of unreliable links \\
      }

\author{
\authorblockN{Osman Ya\u{g}an and Armand M. Makowski}
\authorblockA{Department of Electrical and Computer Engineering\\
              and the Institute for Systems Research\\
              University of Maryland, College Park\\
              College Park, Maryland 20742\\
              oyagan@umd.edu, armand@isr.umd.edu}
}

\maketitle

\begin{abstract}
\normalsize We investigate the secure connectivity of wireless
sensor networks under the pairwise key distribution scheme of Chan
et al.. Unlike recent work which was carried out under the
assumption of {\em full visibility}, here we assume a (simplified)
communication model where unreliable wireless links are
represented as on/off channels. We present conditions on how to
scale the model parameters so that the network i) has no secure
node which is isolated and ii) is securely connected, both with
high probability when the number of sensor nodes becomes large.
The results are given in the form of {\em zero-one laws}, and
exhibit significant differences with corresponding results in the
full visibility case. Through simulations these zero-one laws 
are shown to be valid also under a more
realistic communication model, i.e., the disk model.
\end{abstract}


\section{Introduction}
\label{sec:Introduction}

Wireless sensor networks (WSNs) are distributed collections of
sensors with limited capabilities for computations and wireless
communications. It is envisioned \cite{Akyildiz} that WSNs will be
used in a wide range of applications areas such as healthcare
(e.g.  patient monitoring), military operations (e.g., battlefield
surveillance) and homes (e.g., home automation and monitoring).
These WSNs will often be deployed in hostile environments where
communications can be monitored, and nodes are subject to capture
and surreptitious use by an adversary. Under such circumstances,
cryptographic protection will be needed to ensure secure
communications, and to support functions such as sensor-capture
detection, key revocation and sensor disabling.

Unfortunately, many security schemes developed for general network
environments do not take into account the unique features of WSNs:
Public key cryptography is not feasible computationally because of
the severe limitations imposed on the physical memory and power
consumption of the individual sensors. Traditional key exchange
and distribution protocols are based on trusting third parties,
and this makes them inadequate for large-scale WSNs whose
topologies are unknown prior to deployment. We refer the reader to
the papers
\cite{CamtepeYener,EschenauerGligor,PerrigStankovicWagner} for
discussions of the security challenges in WSN settings.

{\em Random} key predistribution schemes were introduced to
address some of these difficulties. The idea of randomly assigning
secure keys to sensor nodes prior to network deployment was first
introduced by Eschenauer and Gligor \cite{EschenauerGligor}. Since
then, many competing alternatives to the Eschenauer and Gligor
(EG) scheme have been proposed; see \cite{CamtepeYener} for a
detailed survey of various key distribution schemes for WSNs. With
so many schemes available, a basic question arises as to how they
compare with each other. Answering this question passes through a
good understanding of the properties and performance of the
schemes under consideration, and this can be  achieved in a number
of ways. The approach we use here considers random graph models
naturally induced by a given scheme, and then develops the
scaling laws corresponding to desirable network properties, e.g.,
absence of secure nodes which are isolated, secure connectivity,
etc. This is done with the aim of deriving guidelines 
to {\em dimension} the scheme, namely adjust its parameters so
that these properties occur with high probability as the number of
nodes becomes large.

To date, most of the efforts along these lines have been carried
out under the assumption of {\em full visibility} according to
which sensor nodes are all within communication range of each
other; more on this later: Under this assumption, the EG scheme
gives rise to a class of random graphs known as random key graphs;
relevant results are available in the references
\cite{BlackburnGerke,
DiPietroManciniMeiPanconesiRadhakrishnan2008, EschenauerGligor,
Rybarczyk2009, YaganMakowskiConnectivity}. The q-composite scheme
\cite{ChanPerrigSong}, a simple variation of the EG scheme, was
investigated by Bloznelis et al. \cite{BloznelisJaworskiRybarczyk}
through an appropriate extension of the random key graph model.
Recently, Ya\u{g}an and Makowski have analyzed various random
graphs induced by the random pairwise key predistribution scheme
of Chan et al. \cite{ChanPerrigSong}; see the conference papers
\cite{YaganMakowskiPairwise2010, YaganMakowskiGradual2010}.

To be sure, the full visibility assumption does away with the
wireless nature of the communication medium supporting WSNs. In
return, this simplification makes it possible to focus on how
randomization of the key distribution mechanism alone affects the
establishment of a secure network in the best of circumstances,
i..e., when there are no link failures. 
A common criticism of this line of work is that by disregarding the
unreliability of the wireless links, the resulting dimensioning
guidelines are likely to be too {\em optimistic}: In practice
nodes will have fewer neighbors since some of the communication
links may be impaired. As a result, the desired connectivity
properties may not be achieved if dimensioning is done according
to results derived under full visibility.

In this paper, in an attempt to go beyond full visibility, we
revisit the pairwise key predistribution scheme of Chan et al.
\cite{ChanPerrigSong} under more realistic assumptions that
account for the possibility that communication links between nodes
may not be available -- This could occur due to the presence of
physical barriers between nodes or because of harsh environmental
conditions severely impairing transmission. To study such
situations, we introduce a simple communication model where
channels are mutually independent, and are either on or off. 
An overall system model is then constructed by {\em
intersecting} the random graph model of the pairwise key
distribution scheme (under full visibility), with an
Erd\H{o}s-R\'enyi (ER) graph model \cite{Bollobas}. For this new
random graph structure, we establish zero-one laws for two basic
(and related) graph properties, namely graph connectivity and the
absence of isolated nodes, as the model parameters are scaled
with the number of users -- We identify the critical thresholds and
show that they coincide. To the best of our knowledge,
these full zero-one laws 
constitute the first {\em complete} analysis 
of a key distribution scheme under {\em non}-full visibility --
Contrast this with the partial results 
by Yi et al. \cite{YiWanLinHuang} for the absence of
isolated nodes (under additional conditions)
when the communication model is the disk model.

Although the communication model considered here may be deemed
simplistic, it does permit a complete analysis of the issues of
interest, with the results already yielding a number of
interesting observations: 
The obtained zero-one laws differ significantly 
from the corresponding results
in the full visibility case \cite{YaganMakowskiPairwise2010}.
Thus, the communication model may have a
significant impact on the dimensioning of the pairwise
distribution algorithm, and this points to the need of possibly
reevaluating guidelines developed under the full visibility
assumption. Furthermore, simulations suggest that the zero-one
laws obtained here for the on/off channel model may still be
useful in dimensioning the pairwise scheme under the popular, and
more realistic, disk model \cite{GuptaKumar}.

We also compare the results established here with well-known
zero-one laws for ER graphs \cite{Bollobas}. In particular, we
show that the connectivity behavior of the model studied here does
not in general resemble that of the ER graphs. 
The picture is somewhat more subtle for
the results also imply that if the channel is very poor,
the model studied here indeed behaves like an ER graph as
far as connectivity is concerned.
The comparison with ER graphs
is particularly relevant to the analysis of
key distribution schemes for WSNs: 
Indeed, 
connectivity results for ER graphs have often been used in the
dimensioning and evaluation of key distribution schemes, e.g., see
the papers by Eschenauer and Gligor \cite{EschenauerGligor},
Chan et al. \cite{ChanPerrigSong} and Hwang and Kim \cite{HwangKim}. 
There it is a common practice to assume that
the random graph induced by the particular key distribution scheme
behaves {\em like} an ER graph (although it is not strictly
speaking an ER graph). As pointed out by Di Pietro et al.
\cite{DiPietroManciniMeiPanconesiRadhakrishnan2008} such an
assumption is made without any formal justification, and
subsequent efforts to confirm its validity have remained limited
to this date: The EG scheme has been analyzed by a number of
authors \cite{BlackburnGerke,
DiPietroManciniMeiPanconesiRadhakrishnan2008,Rybarczyk2009,
YaganMakowskiConnectivity}, and as a result of these efforts it is
now known that the ER {\em assumption} does yield the correct
results for both the absence of isolated nodes and connectivity
under the assumption of full visibility. 
On the other hand the recent paper \cite{YaganMakowskiPairwise2010}
shows that the ER assumption is not valid for the pairwise key distribution
of Chan et al. \cite{ChanPerrigSong}; 
see Section \ref{subsec:FullVisibility} for details.

The rest of the paper is organized as follows: In Section
\ref{sec:ImplementingPairwise}, we give precise definitions and
implementation details of the pairwise scheme of Chan et al. while
Section \ref{sec:Model} is devoted to describing the model of
interest. The main results of the paper, namely Theorem
\ref{thm:OneLaw+NodeIsolation} and Theorem
\ref{thm:OneLaw+Connectivity}, are presented in Section
\ref{sec:MainResults} with an extensive discussion given in
Section \ref{sec:Remarks}. The remaining sections, namely Sections
\ref{sec:ProofTheoremNodeIsolation} through
\ref{sec:ProofConnectivityII}, are devoted to establishing the
main results of the paper.

A word on notation and conventions in use: All limiting
statements, including asymptotic equivalences, are understood with
$n$ going to infinity. The random variables (rvs) under
consideration are all defined on the same probability triple
$(\Omega, {\cal F}, \mathbb{P})$. Probabilistic statements are
made with respect to this probability measure $\mathbb{P}$, and we
denote the corresponding expectation operator by $\mathbb{E}$.
Also, we use the notation $=_{st}$ to indicate distributional
equality. The indicator function of an event $E$ is denoted by
$\1{E}$. For any discrete set $S$ we write $|S|$ for its
cardinality. Also, for any pair of events $E$ and $F$ we have
\begin{equation}
\1{ E \cup F } = \1{E} + \1{F} - \1{E \cap F}.
\label{eq:BasicSetIdentity}
\end{equation}

\section{Implementing pairwise key distribution schemes}
\label{sec:ImplementingPairwise}

Interest in the random pairwise key predistribution scheme of Chan
et al. \cite{ChanPerrigSong} stems from the following advantages
over the EG scheme: (i) Even if some nodes are captured, the
secrecy of the remaining nodes is {\em perfectly} preserved; (ii)
Unlike earlier schemes, this pairwise scheme enables both
node-to-node authentication and quorum-based revocation.

As in the conference papers \cite{YaganMakowskiPairwise2010,
YaganMakowskiGradual2010}, we parametrize the pairwise key
distribution scheme by two positive integers $n$ and $K$ such that
$K < n$. There are $n$ nodes, labelled $i=1, \ldots , n$, with
unique ids ${\rm Id}_1, \ldots , {\rm Id}_n$. Write ${\cal N} :=
\{ 1, \ldots n \}$ and set ${\cal N}_{-i} := {\cal N}-\{i\}$ for
each $i=1, \ldots , n$. With node $i$ we associate a subset
$\Gamma_{n,i}$ of nodes selected at {\em random} from ${\cal
N}_{-i}$ -- We say that each of the nodes in $\Gamma_{n,i}$ is
paired to node $i$. Thus, for any subset $A \subseteq {\cal
N}_{-i}$, we require
\[
\bP{ \Gamma_{n,i} = A } = \left \{
\begin{array}{ll}
{{n-1}\choose{K}}^{-1} & \mbox{if $|A|=K$} \\
              &                   \\
0             & \mbox{otherwise.}
\end{array}
\right .
\]
The selection of $\Gamma_{n,i}$ is done {\em uniformly} amongst
all subsets of ${\cal N}_{-i}$ which are of size exactly $K$. The
rvs $\Gamma_{n,1}, \ldots , \Gamma_{n,n}$ are assumed to be
mutually independent so that
\[
\bP{ \Gamma_{n,i} = A_i, \ i=1, \ldots , n } = \prod_{i=1}^n \bP{
\Gamma_{n,i} = A_i }
\]
for arbitrary $A_1, \ldots , A_n$ subsets of ${\cal N}_{-1},
\ldots , {\cal N}_{-n} $, respectively.

Once this {\em offline} random pairing has been created, we
construct the key rings $\Sigma_{n,1}, \ldots , \Sigma_{n,n}$, one
for each node, as follows: Assumed available is a collection of
$nK$ distinct cryptographic keys $\{ \omega_{i|\ell}, \ i=1,
\ldots , n ; \ \ell=1, \ldots , K \}$. Fix $i=1, \ldots , n$ and
let $\ell_{n,i}: \Gamma_{n,i} \rightarrow \{ 1, \ldots , K \}$
denote a labeling of $\Gamma_{n,i}$. For each node $j$ in
$\Gamma_{n,i}$ paired to $i$, the cryptographic key
$\omega_{i|\ell_{n,i}(j)}$ is associated with $j$. For instance,
if the random set $\Gamma_{n,i}$ is realized as $\{ j_1, \ldots ,
j_K \}$ with $1 \leq j_1 < \ldots < j_K \leq n $, then an obvious
labeling consists in $\ell_{n,i}(j_k) = k $ for each $k=1, \ldots
, K$ with key $\omega_{i|k}$ associated with node $j_k$. Of course
other labeling are possible, e.g., according to decreasing labels
or according to a random permutation. Finally, the pairwise key $
\omega^\star_{n,ij} = [ {\rm Id}_i | {\rm Id}_j |
\omega_{i|\ell_{n,i}(j)} ] $ is constructed and inserted in the
memory modules of both nodes $i$ and $j$. The key
$\omega^\star_{n,ij}$ is assigned {\em exclusively} to the pair of
nodes $i$ and $j$, hence the terminology pairwise distribution
scheme. The key ring $\Sigma_{n,i}$ of node $i$ is the set
\begin{equation}
\Sigma_{n,i} := \{ \omega^\star_{n,ij}, \ j \in \Gamma_{n,i} \}
\cup \{ \omega^\star_{n,ji}, \ i \in \Gamma_{n,j} \}.
\label{eq:KeyRingDefn}
\end{equation}

If two nodes, say $i$ and $j$, are within communication range of
each other, then they can establish a secure link if at least one
of the events $i \in \Gamma_{n,j}$ or $j \in \Gamma_{n,j}$ is
taking place. Both events can take place, in which case the memory
modules of node $i$ and $j$ both contain the distinct keys
$\omega^\star_{n,ij}$ and $\omega^\star_{n,ji}$. Finally, it is
plain by construction that this scheme supports node-to-node
authentication.

\section{The model}
\label{sec:Model}

Under full visibility, this pairwise distribution scheme naturally
gives rise to the following class of random graphs: With $n=2,3,
\ldots $ and positive integer $K < n$, we say that the distinct
nodes $i$ and $j$ are K-adjacent, written $i \sim_K j$, if and
only if they have at least one key in common in their key rings,
namely
\begin{equation}
i \sim_K j \quad \mbox{iff} \quad \Sigma_{n,i} \cap \Sigma_{n,j}
\neq \emptyset . \label{eq:Adjacency}
\end{equation}
Let $\mathbb{H}(n;K)$ denote the undirected random graph on the
vertex set $\{ 1, \ldots , n \}$ induced by the adjacency notion
(\ref{eq:Adjacency}); this corresponds to modelling the pairwise
distribution scheme under full visibility. We have
\begin{equation}
\bP{i \sim_{K} j } = \lambda_n (K) \label{eq:edge_prob_key_graph}
\end{equation}
where $\lambda_n(K)$ is the link assignment probability in
$\mathbb{H}(n;K)$ given by
\begin{eqnarray}
\lambda_n (K) &=& 1 - \left ( 1 - \frac{K}{n-1} \right )^2
\nonumber \\
&=& \frac{2K}{n-1}-\left(\frac{K}{n-1}\right)^2 .
\label{eq:LinkAssignmentinH}
\end{eqnarray}

As mentioned earlier, in this paper we seek to account for the
possibility that communication links between nodes may not be
available. To study such situations, we assume a communication
model that consists of independent channels each of which can be
either on or off. Thus, with $p$ in $(0,1)$, let $\{B_{ij}(p), 1
\leq i < j \leq n\}$ denote i.i.d. $\{0, 1\}$-valued rvs with
success probability $p$. The channel between nodes $i$ and $j$ is
available (resp. up) with probability $p$ and unavailable (resp.
down) with the complementary probability $1-p$.

Distinct nodes $i$ and $j$ are said to be B-adjacent, written $i
\sim_{B} j$, if $B_{ij}(p) = 1$. The notion of B-adjacency defines
the standard ER graph $\mathbb{G}(n;p)$ on the vertex set $\{ 1,
\ldots , n \}$. Obviously,
\[
\bP{ i \sim j}_{B} = p.
\]

The random graph model studied here is obtained by {\em
intersecting} the random pairwise graph $\mathbb{H}(n;K)$ with the
ER graph $\mathbb{G}(n;p)$. More precisely, the distinct nodes $i$
and $j$ are said to be adjacent, written $i \sim j$, if and only
they are both K-adjacent and B-adjacent, namely
\begin{equation}
i \sim j \quad \mbox{iff} \quad
\begin{array}{c}
\Sigma_{n,i} \cap \Sigma_{n,j} \neq \emptyset \\
\mbox{and} \\
B_{ij}(p)=1.\\
\end{array}
\label{eq:Adjacency_Intersection}
\end{equation}
The resulting {\em undirected} random graph defined on the vertex
set $\{1, \ldots, n\}$ through this notion of adjacency is denoted
$\mathbb{H\cap G}(n;K,p)$.

Throughout the collections of rvs $\{ \Gamma_{n,1}, \ldots ,
\Gamma_{n,n} \}$ and $\{B_{ij}(p), 1 \leq i < j \leq n\}$ are
assumed to be independent, in which case the edge occurrence
probability in $\mathbb{H\cap G}(n;K,p)$ is given by
\begin{equation}
\bP{i \sim j } = p \cdot \bP{i \sim_{K} j } = p \lambda_n (K) .
\label{eq:edge_prob_intersectioN_graph}
\end{equation}

\section{The results}
\label{sec:MainResults}

To fix the terminology, we refer to any mapping $K: \mathbb{N}_0
\rightarrow \mathbb{N}_0$ as a {\em scaling} (for random pairwise
graphs) provided it satisfies the natural conditions
\begin{equation}
K_n < n, \quad n=1,2, \ldots . \label{eq:ScalingDefn}
\end{equation}
Similarly, any mapping $p: \mathbb{N}_0 \rightarrow (0,1)$ defines
a scaling for ER graphs.

To lighten the notation we often group the parameters $K$ and $p$
into the ordered pair $\theta \equiv (K,p)$. Hence, a mapping
$\theta: \mathbb{N}_0 \rightarrow \mathbb{N}_0 \times (0,1)$
defines a scaling for the intersection graph $\mathbb{H\cap
G}(n;\theta)$ provided the condition (\ref{eq:ScalingDefn}) holds
on the first component.

The results will be expressed in terms of the threshold function
$\tau: [0,1] \rightarrow [0,1]$ defined by
\begin{equation}
\tau(p)=\left \{
\begin{array}{lll}
1 & \mbox{if~ $p=0$} \\
  &                      \\
\frac{2}{1-\frac{\log(1-p)}{p}} & \mbox{if~$0 < p < 1$} \\
 &                      \\
0 & \mbox{if~$p = 1$.}
\end{array}
\right . \label{eq:threshold}
\end{equation}
It is easy to check that this threshold function is continuous on
its entire domain of definition; see Figure \ref{figure1}.

\subsection{Absence of isolated nodes}
\label{subsec:ResultsAbsenceIsolatedNodes}

The first result gives a zero-one law for the absence of isolated
nodes.

\begin{theorem}
{\sl Consider scalings $K: \mathbb{N}_0 \rightarrow \mathbb{N}_0$
and $p: \mathbb{N}_0 \rightarrow (0,1)$ such that
\begin{equation}
p_n\left(2K_n-\frac{K_n ^ 2}{n-1}\right) \sim c \log n, \quad
n=1,2, \ldots \label{eq:scalinglaw}
\end{equation}
for some $c>0$. If $\lim_{n \to \infty}p_n=p^\star$ for some
$p^\star$ in $[0,1]$, then we have
\begin{eqnarray}
\lefteqn{ \lim_{n \rightarrow \infty } \bP{
\begin{array}{c}
\mathbb{H \cap G}(n;\theta_n)~\mbox{contains} \\
\mbox{~no~isolated~nodes} \\
\end{array}
} } & &
\nonumber \\
&=& \left \{
\begin{array}{ll}
0 & \mbox{if~ $c < \tau(p^\star)$} \\
  &                      \\
1 & \mbox{if~$c > \tau(p^\star)$.}
\end{array}
\right . \label{eq:OneLaw+NodeIsolation}
\end{eqnarray}
} \label{thm:OneLaw+NodeIsolation}
\end{theorem}

The condition (\ref{eq:scalinglaw}) on the scaling $\mathbb{N}_0
\rightarrow (0,1) \times \mathbb{N}_0$ will often be used in the
equivalent form
\begin{equation}
p_n\left(2K_n-\frac{K_n ^ 2}{n-1}\right) = c_n \log n, \quad
n=1,2, \ldots \label{eq:scalinglawEquivalent}
\end{equation}
with the sequence $c: \mathbb{N}_0 \rightarrow \mathbb{R}_+$
satisfying $\lim_{n \rightarrow \infty} c_n = c$.

\subsection{Connectivity}
\label{subsec:ResultsConnectivity}

An analog of Theorem \ref{thm:OneLaw+NodeIsolation} also holds for
the property of graph connectivity.

\begin{theorem}
{\sl Consider scalings $K: \mathbb{N}_0 \rightarrow \mathbb{N}_0$
and $p: \mathbb{N}_0 \rightarrow (0,1)$ such that
(\ref{eq:scalinglaw}) holds for some $c>0$. If $\lim_{n \to
\infty}p_n=p^\star$ for some $p^\star$ in $[0,1]$, then we have
\begin{eqnarray}
\lefteqn{ \lim_{n \rightarrow \infty } \bP{\mathbb{H \cap
G}(n;\theta_n)~\mbox{ is connected} } } & &
\nonumber \\
&=& \left \{
\begin{array}{ll}
0 & \mbox{if~ $c < \tau(p^\star)$} \\
  &                      \\
1 & \mbox{if~$c > \tau(p^\star)$}
\end{array}
\right . \label{eq:OneLaw+Connectivity}
\end{eqnarray}
where the threshold $\tau(p^\star)$ is given by
(\ref{eq:threshold}). } \label{thm:OneLaw+Connectivity}
\end{theorem}

Comparing Theorem \ref{thm:OneLaw+Connectivity} with Theorem
\ref{thm:OneLaw+NodeIsolation}, we see that the class of random
graphs studied here provides one more instance where the zero-one
laws for absence of isolated nodes and connectivity coincide, viz.
ER graphs \cite{Bollobas}, random geometric graphs
\cite{PenroseBook} or random key graphs
\cite{BlackburnGerke,Rybarczyk2009,YaganMakowskiConnectivity}.

A case of particular interest arises when $p^\star > 0$ since
requiring (\ref{eq:scalinglaw}) now amounts to
\begin{equation}
\left(2K_n-\frac{K_n ^ 2}{n-1}\right) \sim ~ \frac{c}{p^\star}
\log n \label{eq:scalinglawEquivalent2}
\end{equation}
for some $c>0$. Any scaling $K: \mathbb{N}_0 \rightarrow
\mathbb{N}_0$ which behaves like (\ref{eq:scalinglawEquivalent2})
must necessarily satisfy $K_n = o(n)$, and it is easy to see that
requiring (\ref{eq:scalinglaw}) is equivalent to
\begin{equation}
K_n \sim ~ t \log  n \label{eq:scalinglawEquivalent3}
\end{equation}
for some $t>0$ with $c$ and $t$ related by $t=\frac{c}{2p^\star}$.
With this reparametrization, Theorem
\ref{thm:OneLaw+NodeIsolation} and Theorem
\ref{thm:OneLaw+Connectivity} can be summarized in the following
simpler form:

\begin{theorem}
{\sl Consider scalings $K: \mathbb{N}_0 \rightarrow \mathbb{N}_0$
and $p: \mathbb{N}_0 \rightarrow (0,1)$ such that $\lim_{n \to
\infty}p_n=p^\star > 0$. Under the condition
(\ref{eq:scalinglawEquivalent3}) for some $t>0$, we have
\begin{eqnarray}
\lefteqn{ \lim_{n \rightarrow \infty } \bP{ \mathbb{H \cap
G}(n;\theta_n)~\mbox{contains~no~isolated~nodes} } } &&
\nonumber \\
&=& \lim_{n \rightarrow \infty } \bP{\mathbb{H \cap
G}(n;\theta_n)~\mbox{ is connected} }
\nonumber \\
&=& \left \{
\begin{array}{ll}
0 & \mbox{if~ $t < \widehat \tau(p^\star)$} \\
  &                      \\
1 & \mbox{if~ $t > \widehat \tau(p^\star)$}
\end{array}
\right . \label{eq:OneLawModified}
\end{eqnarray}
where we have set
\begin{equation}
\widehat \tau (p) := \frac{\tau(p)}{2 p} = \frac{ 1 }{p - \log
(1-p) }, \quad 0 < p < 1. \label{eq:threshold2}
\end{equation}
} \label{thm:OneLawModified}
\end{theorem}

This alternate formulation is particularly relevant for the case
$p_n = p^\star$ (in $(0,1)$) for all $n=1,2, \ldots$, which
captures situations when channel conditions are not affected by
the number of users. Such simplifications do not occur in the more
realistic case $p^\star = 0$ which corresponds to the situation
where channel conditions are indeed influenced by the number of
users in the system -- The more users in the network, the more
likely they will experience interferences from other users.

\begin{figure}[!t]
 \hspace{-0.45 cm}
\includegraphics[totalheight=0.37\textheight,
width=0.55\textwidth]{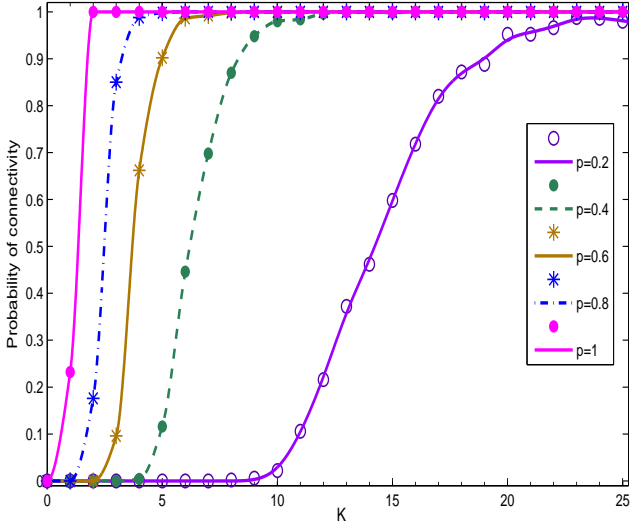} \caption{Probability that
$\mathbb{H} \cap \mathbb{G}(n;K,p)$
         is connected as a function of $K$ for
         $p=0.2$, $p=0.4$, $p=0.6$, $p=0.8$ and $p=1$
         with $n=200$.
} \label{figure:theorem_connect}
\end{figure}

\begin{figure}[t]
 \hspace{-0.45 cm}
\includegraphics[totalheight=0.37\textheight,
width=0.55\textwidth]{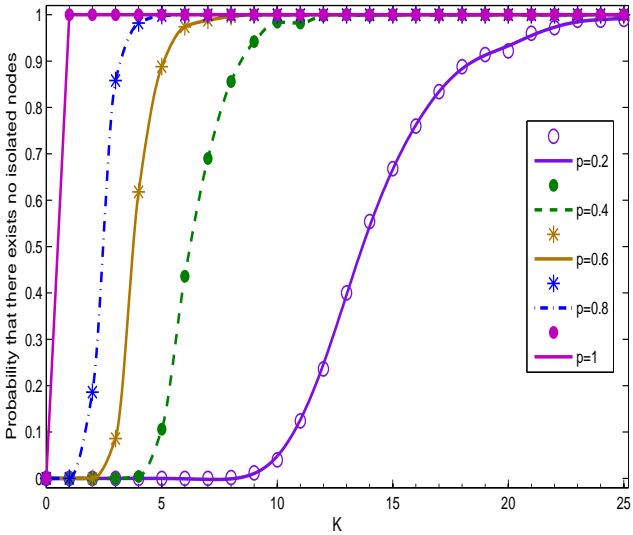} \caption{Probability that
$\mathbb{H} \cap \mathbb{G}(n;K,p)$
         has no isolated nodes as a function of $K$ for
         $p=0.2$, $p=0.4$, $p=0.6$, $p=0.8$ and $p=1$
         with $n=200$.
         This figure clearly resembles
         Figure \ref{figure:theorem_connect} for all $p\neq 1$.
} \label{figure:theorem_isol}
\end{figure}

We now present numerical results that verify
(\ref{eq:OneLawModified}). In all the simulations, we fix the
number of nodes at $n=200$. We consider the channel parameters
$p=0.2$, $p=0.4$, $p=0.6$, $p=0.8$, and $p=1$ (the full visibility
case), while varying the parameter $K$ from $1$ to $25$. For each
parameter pair $(K,p)$, we generate $500$ independent samples of
the graph $\mathbb{H} \cap \mathbb{G}(n;K,p)$ and count the number
of times (out of a possible 500) that the obtained graphs i) have
no isolated nodes and ii) are connected. Dividing the counts by
$500$, we obtain the (empirical) probabilities for the events of
interest. The results for connectivity are depicted in Figure
\ref{figure:theorem_connect}, where the curve fitting tool of
MATLAB is used. It is easy to check that for each value of $p \neq
1$, the connectivity threshold matches the prescription
(\ref{eq:OneLawModified}), namely $K= \widehat \tau (p) \log n$.
It is also seen that, if the channel is poor, i.e., if $p$ is
close to zero, then the required value for $K$ to ensure
connectivity can be much larger than the one in the full
visibility case $p=1$. The results regarding the absence of node
isolation are depicted in Figure \ref{figure:theorem_isol}. For
each value of $p\neq 1$, Figure \ref{figure:theorem_isol} is
indistinguishable from Figure \ref{figure:theorem_connect}, with
the difference between the estimated probabilities of graph
connectivity and absence of isolated nodes being quite small, in
agreement with (\ref{eq:OneLawModified}).

\section{Discussion and comments}
\label{sec:Remarks}

\subsection{Comparing with the full-visibility case}
\label{subsec:FullVisibility}

At this point the reader may wonder as to what form would Theorem
\ref{thm:OneLaw+Connectivity} take in the context of full
visibility-- In the setting developed here this corresponds to $p
= 1$ so that $\mathbb{H \cap G} (n;\theta)$ coincides with
$\mathbb{H} (n;K)$; see the curve for $p=1$ in Figure
\ref{figure:theorem_connect}). Relevant results for this case were
obtained recently by the authors in
\cite{YaganMakowskiPairwise2010}.

\begin{theorem}
{\sl For any $K$ a positive integer, it holds that
\[
\lim_{n \rightarrow \infty } \bP{\mathbb{H}(n;K)~\mbox{ is
connected} } = \left \{
\begin{array}{ll}
0 & \mbox{if~ $K = 1$} \\
  &                      \\
1 & \mbox{if~$K \geq 2$.}
\end{array}
\right . \label{eq:OneLaw+ForConnectivity}
\]
\label{thm:OneLaw+ForConnectivity} }
\end{theorem}

The case where the parameter $K$ is scaled with $n$ is an easy
corollary of Theorem \ref{thm:OneLaw+ForConnectivity}.

\begin{corollary}
{\sl For any scaling $K: \mathbb{N}_0 \rightarrow \mathbb{N}_0$
such that $K_n \geq 2$ for all $n$ sufficiently large, we have the
one-law
\[
\lim_{n \rightarrow \infty } \bP{\mathbb{H}(n;K_n)~\mbox{ is
connected} } = 1.
\]
} \label{cor:OneLaw+ConnectivityUnderScaling}
\end{corollary}
Each node in $\mathbb{H} (n; K)$ has degree at least $K$, so that
no node is ever isolated in $\mathbb{H} (n; K)$. This is in sharp
contrast with the model studied here, as reflected by the full
zero-one law for node isolation given in Theorem
\ref{thm:OneLaw+NodeIsolation}.

Theorem \ref{thm:OneLaw+ForConnectivity} and its Corollary
\ref{cor:OneLaw+ConnectivityUnderScaling} together show that very
small values of $K$ suffice to ensure asymptotically almost sure
(a.a.s.) connectivity of the random graph $\mathbb{H}(n;K)$.
However, these two results cannot be recovered from Theorem
\ref{thm:OneLaw+Connectivity} whose zero-one laws are derived
under the assumption $p_n < 1 $ for all $n=1,2, \ldots$.
Furthermore, even if the scaling $p: \mathbb{N}_0 \rightarrow
(0,1)$ were to satisfy $\lim_{n \rightarrow \infty} p_n = 1$, only
the one-laws in Theorem \ref{thm:OneLawModified} remain since
$\tau(p^\star)=0$ (and $\widehat \tau (p^\star) = 0$) at
$p^\star=1$. Although this might perhaps be expected given the
aforementioned absence of isolated nodes in $\mathbb{H}(n;K)$, the
one-laws for both the absence of isolated nodes and graph
connectivity in $\mathbb{H} \cap \mathbb{G}(n;\theta)$ still
require conditions on the behavior of the scaling $K: \mathbb{N}_0
\rightarrow \mathbb{N}_0$, namely (\ref{eq:scalinglawEquivalent3})
(whereas Corollary \ref{cor:OneLaw+ConnectivityUnderScaling} does
not).

\subsection{Comparing $\mathbb{H\cap G}(n;\theta)$ with ER graphs}

In the original paper of Chan et al. \cite{ChanPerrigSong} (as in
the reference \cite{HwangKim}), the connectivity analysis of the
pairwise scheme was based on ER graphs \cite{Bollobas} -- It was
assumed that the random graph induced by the pairwise scheme under
a communication model (taken mostly to be the disk model
\cite{GuptaKumar}) behaves {\em like} an ER graph; similar
assumptions have been made in \cite{EschenauerGligor, HwangKim}
when discussing the connectivity of the EG scheme. However, this
assumption was made without any formal justification. Recently we
have shown that the full visibility model $\mathbb{H}(n;K)$ has
major differences with an ER graph. For instance, the edge
assignments are (negatively) correlated in $\mathbb{H}(n; K)$
while independent in ER graphs; see
\cite{YaganMakowskiPairwise2010} for a detailed discussion on the
differences of $\mathbb{H}(n;K)$ and $\mathbb{G}(n;p)$. It is easy
to verify that the edge assignments in $\mathbb{H\cap
G}(n;\theta)$ are also negatively correlated; see Section
\ref{sec:NegativeDependence}. Therefore, the models
$\mathbb{H}(n;K)$ and $\mathbb{H\cap G}(n;\theta)$ 
cannot be equated with an ER graph, and the results
obtained in \cite{YaganMakowskiPairwise2010} 
and in this paper are {\em not} mere consequences of
classical results for ER graphs.

\begin{figure}[t]
\hspace{-0.5 cm}
\includegraphics[totalheight=0.37\textheight,
width=0.55\textwidth]{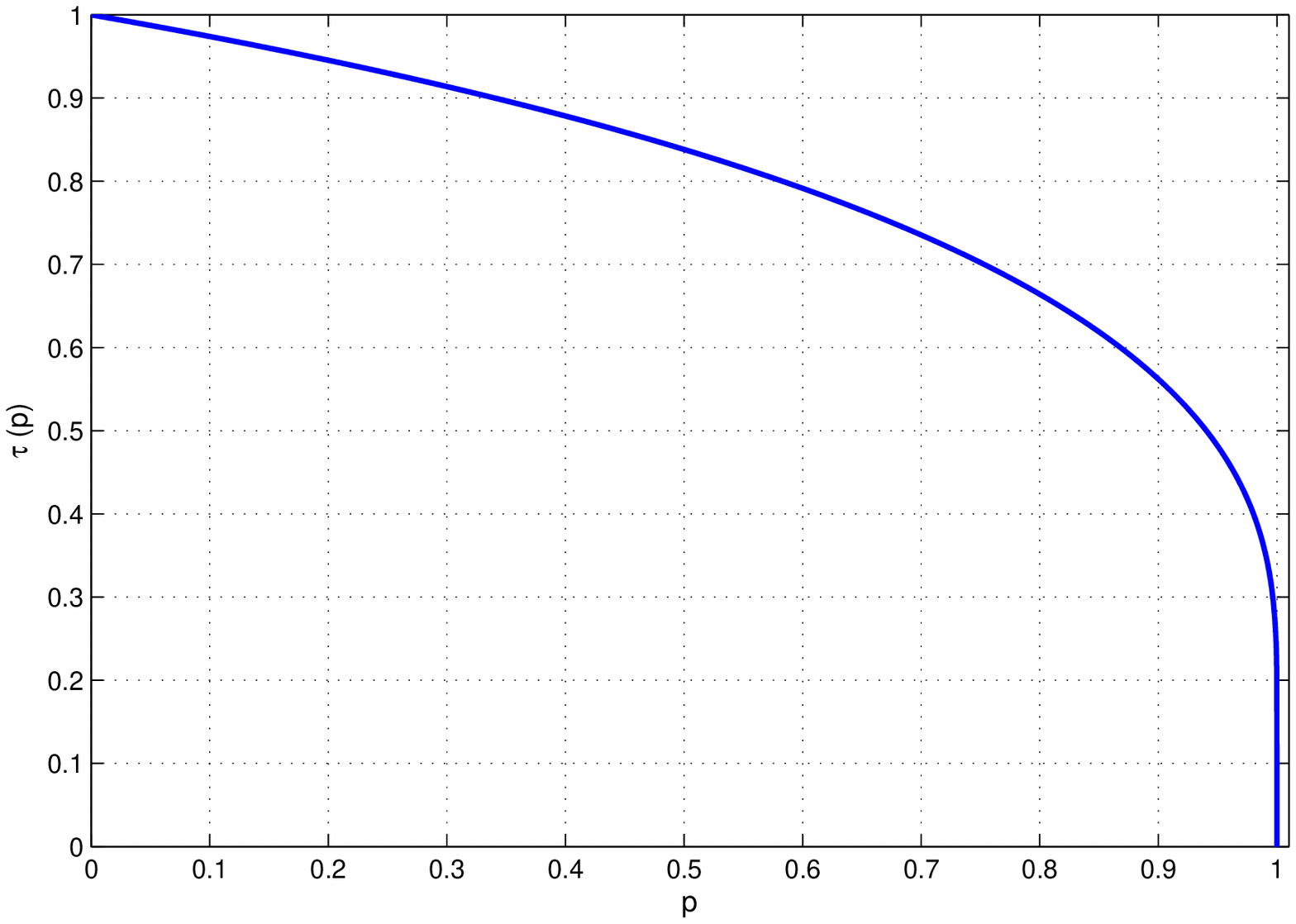} \caption{$\tau(p)$ vs $p$.
         Clearly $\tau(p^ \star)=1$ only if
         $\lim_{n \to \infty} p_n = p^\star = 0$.}
\label{figure1}
\end{figure}

However, {\em formal} similarities do exist between $\mathbb{H\cap
G}(n;\theta)$ and ER graphs. Recall the following well-known
zero-one law for ER graphs: For any scaling $p: \mathbb{N}_0
\rightarrow [0,1]$ satisfying
\[
p_n \sim c ~ \frac{ \log n}{n}
\]
for some $c>0$, it holds that
\[
\lim_{n \rightarrow \infty } \bP{ \mathbb{G}(n;p_n)  ~\mbox{\rm
is~connected} } = \left \{
\begin{array}{ll}
0 & \mbox{if~ $c < 1$} \\
  &                 \\
1 & \mbox{if~ $c > 1 $.}
\end{array}
\right .
\]
On the other hand, the condition (\ref{eq:scalinglaw}) can be
rephrased more compactly as
\[
p_n \lambda_n (K_n) \sim c ~ \frac{\log n }{n}, \qquad c>0
\]
with the results (\ref{eq:OneLaw+NodeIsolation}) and
(\ref{eq:OneLaw+Connectivity}) unchanged. Hence, in both ER graphs
and $\mathbb{H\cap G}(n;\theta)$, the zero-one laws can be
expressed as a comparison of the probability of link assignment
against the critical scaling $\frac{\log n}{n}$; this is also the
case for random geometric graphs \cite{PenroseBook}, and random
key graphs
\cite{BlackburnGerke,Rybarczyk2009,YaganMakowskiConnectivity}. But
the condition $c>\tau(p^ \star)$ that ensures a.a.s. connectivity
in $\mathbb{H} \cap \mathbb{G}(n;\theta)$ is not the same as the
condition $c>1$ for a.a.s. connectivity in ER graphs; see Figure
\ref{figure1}. Thus, the connectivity behavior of the model
$\mathbb{H} \cap \mathbb{G}(n;\theta)$ is in general different
from that in an ER graph, and a \lq \lq transfer" of the
connectivity results from ER graphs cannot be taken for granted.
Yet, the comparison becomes intricate when the channel is poor:
The connectivity behaviors of the two models do match in the
practically relevant case (for WSNs) $\lim_{n \to \infty}p_n = 0$
since $\tau(0)=1$. 

\subsection{A more realistic communication model}

\begin{figure}[t]
 \hspace{-0.45 cm}
\includegraphics[totalheight=0.37\textheight,
width=0.55\textwidth]{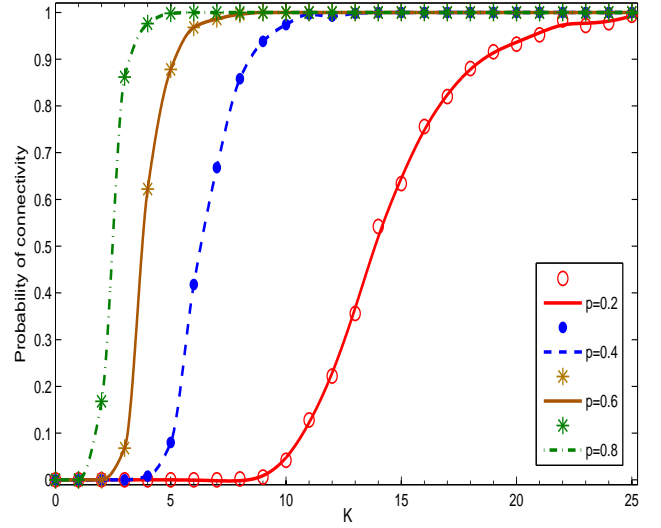} \caption{Probability that
         $\mathbb{H} \cap \mathbb{G}(n;K,\rho)$
         is connected as a function of $K$.
         The number of nodes is set to $n=200$ and $\rho$ is given
         by $\pi \rho ^2 = p$.
} \label{figure:disk}
\end{figure}

One possible extension of the work presented here would be to
consider a more realistic communication model; e.g., the popular
disk model \cite{GuptaKumar} which takes into account the
geographical positions of the sensor nodes. For instance, assume
that the nodes are distributed over a bounded region $\mathcal{D}$
of the plane. According to the {\em disk model}, nodes $i$ and $j$
located at $\boldsymbol{x_i}$ and $\boldsymbol{x_j}$,
respectively, in $\mathcal{D}$ are able to communicate if
\begin{equation}
\parallel \boldsymbol{x_i} -\boldsymbol{x_j} \parallel<\rho
\label{eq:cond_wireless_range}
\end{equation}
where $\rho >0$ is called the transmission range. When the node
locations are independently and randomly distributed over the
region $\mathcal{D}$, the graph induced under the condition
(\ref{eq:cond_wireless_range}) is known as a random geometric
graph \cite{PenroseBook}, thereafter denoted $\mathbb{G}(n;\rho)$.

Under the disk model, studying the pairwise scheme of Chan et al.
amounts to analyzing the intersection of $\mathbb{H}(n;K)$ and
$\mathbb{G}(n;\rho)$, say $\mathbb{H\cap G}(n;K,\rho)$. A direct
analysis of this model seems to be very challenging; see below for
more on this. However, limited simulations already suggest that
the zero-one laws obtained here for $\mathbb{H\cap G}(n;K,p)$ have
an analog for the model $\mathbb{H\cap G}(n;K,\rho)$. To verify
this, consider $200$ nodes distributed uniformly and independently
over a folded unit square $[0,1]^2$ with toroidal (continuous)
boundary conditions. Since there are no border effects, it is easy
to check that
\[
\bP{\: \parallel \boldsymbol{x_i} -\boldsymbol{x_j} \parallel<\rho
\:} = \pi \rho ^2, \quad i \neq j, \:\: i,j=1,2, \ldots, n.
\]
whenever $\rho<0.5$. We match the two communication models
$\mathbb{G}(n;p)$ and $\mathbb{G}(n;\rho)$ by requiring $\pi
\rho^2 = p$. Then, we use the same procedure that produced Figure
\ref{figure:theorem_connect} to obtain the empirical probability
that $\mathbb{H\cap G}(n;K,\rho)$ is connected for various values
of $K$ and $p$. The results are depicted in Figure
\ref{figure:disk} whose resemblance with Figure
\ref{figure:theorem_connect} suggests that the connectivity
behaviors of the models $\mathbb{H\cap G}(n;K,p)$ and
$\mathbb{H\cap G}(n;K,\rho)$ are quite similar. This raises the
possibility that the results obtained here for the on/off
communication model can also be used for dimensioning the pairwise
scheme under the disk model.

A complete analysis of $\mathbb{H\cap G}(n;K,\rho)$ is likely to
be very challenging given the difficulties already encountered in
the analysis of similar problems. For example, the intersection of
random geometric graphs with ER graphs was considered in
\cite{PrasanthMakowksi,YiWanLinHuang}. Although zero-one laws for
graph connectivity are available for each component random graph,
the results for the intersection model in
\cite{PrasanthMakowksi,YiWanLinHuang} were limited only to the
absence of isolated nodes; the connectivity problem is still open
for that model. Yi et al. \cite{YiWanLinHuang} also consider the
intersection of random key graphs with random geometric graphs,
but these results are again limited to the property of node
isolation. To the best of our knowledge, Theorem
\ref{thm:OneLaw+Connectivity} reported here constitutes the only
zero-one law for graph connectivity in a model formed by
intersecting multiple random graphs! (Except of course the trivial
case where an ER graph intersects another ER graph.)

\subsection{Intersection of random graphs}

\begin{figure}[!t]
\hspace{-0.75 cm}
\includegraphics[totalheight=0.37\textheight,
width=0.55\textwidth]{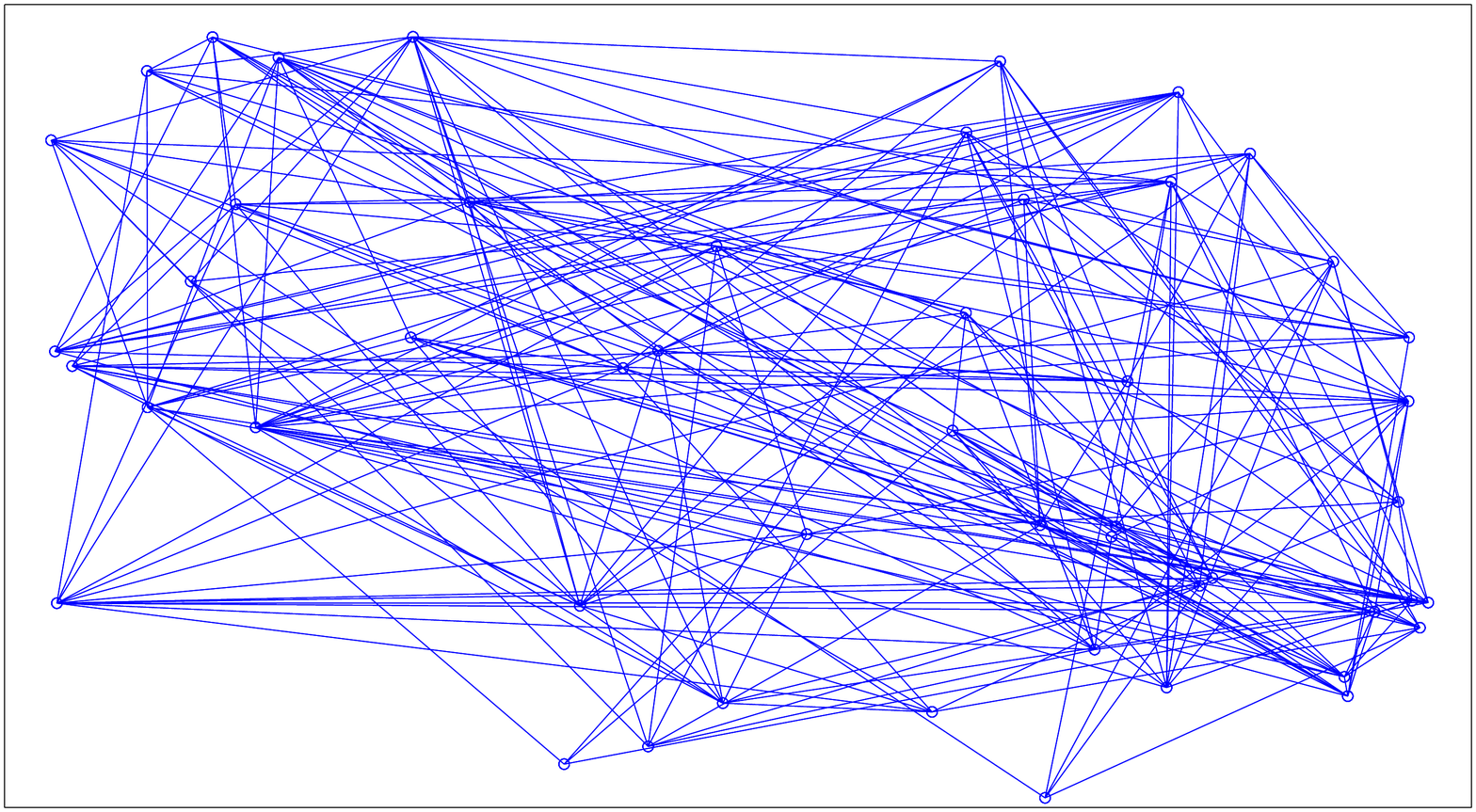} \caption{An instantiation of ER
graph $\mathbb{G}(n;p)$
         with $n=50$ and $p=0.2$.-- The
         graph is connected.}
\label{figure:channel}
\end{figure}

\begin{figure}[!t]
\hspace{-0.55 cm}
\includegraphics[totalheight=0.37\textheight,
width=0.55\textwidth]{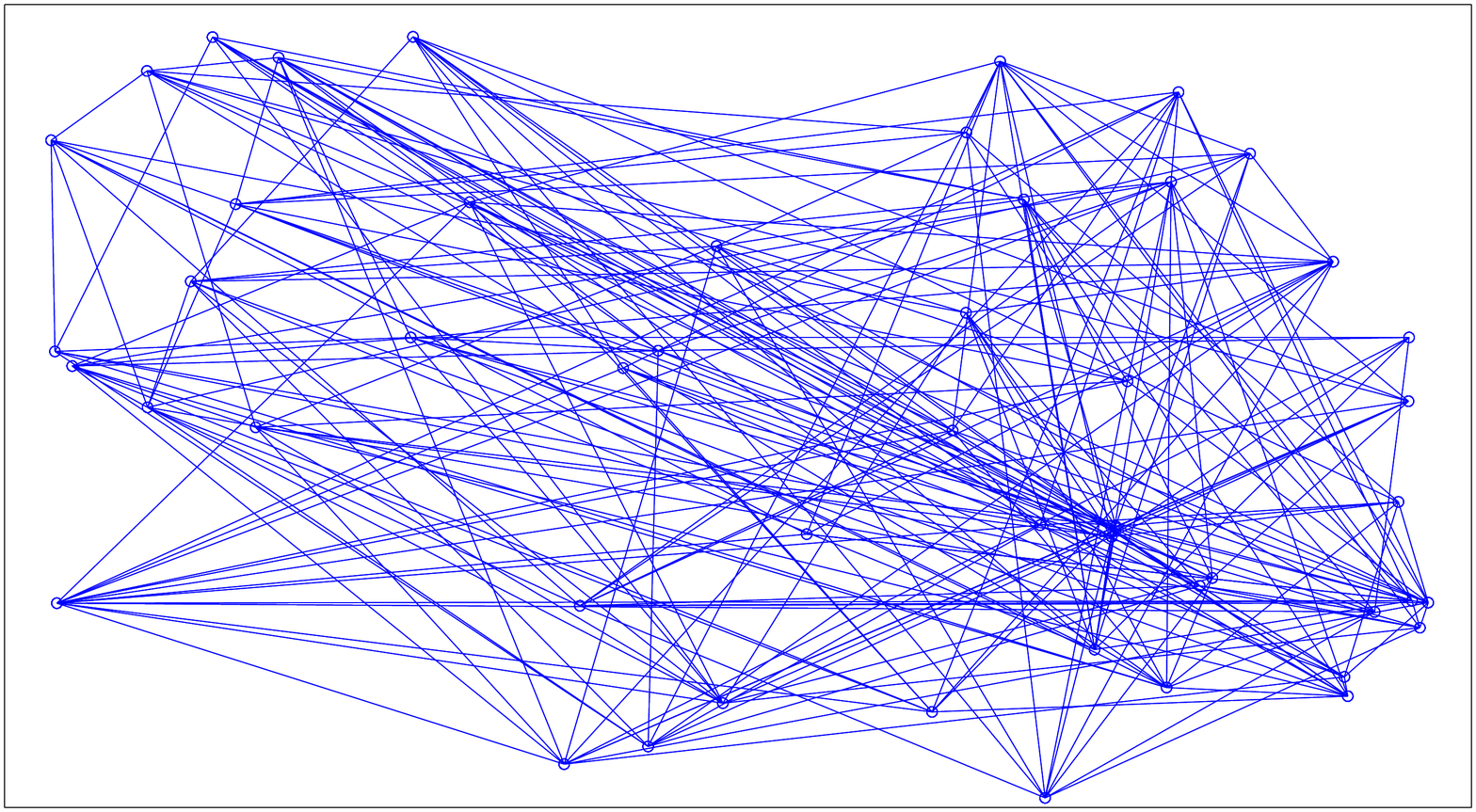} \caption{An instantiation of
$\mathbb{H}(n;K)$
         with $n=50$ and $K=5$.-- The graph is connected.}
\label{figure:pairwise_graph}
\end{figure}

\begin{figure}[!h]
\hspace{-0.5 cm}
\includegraphics[totalheight=0.37\textheight,
width=0.55\textwidth]{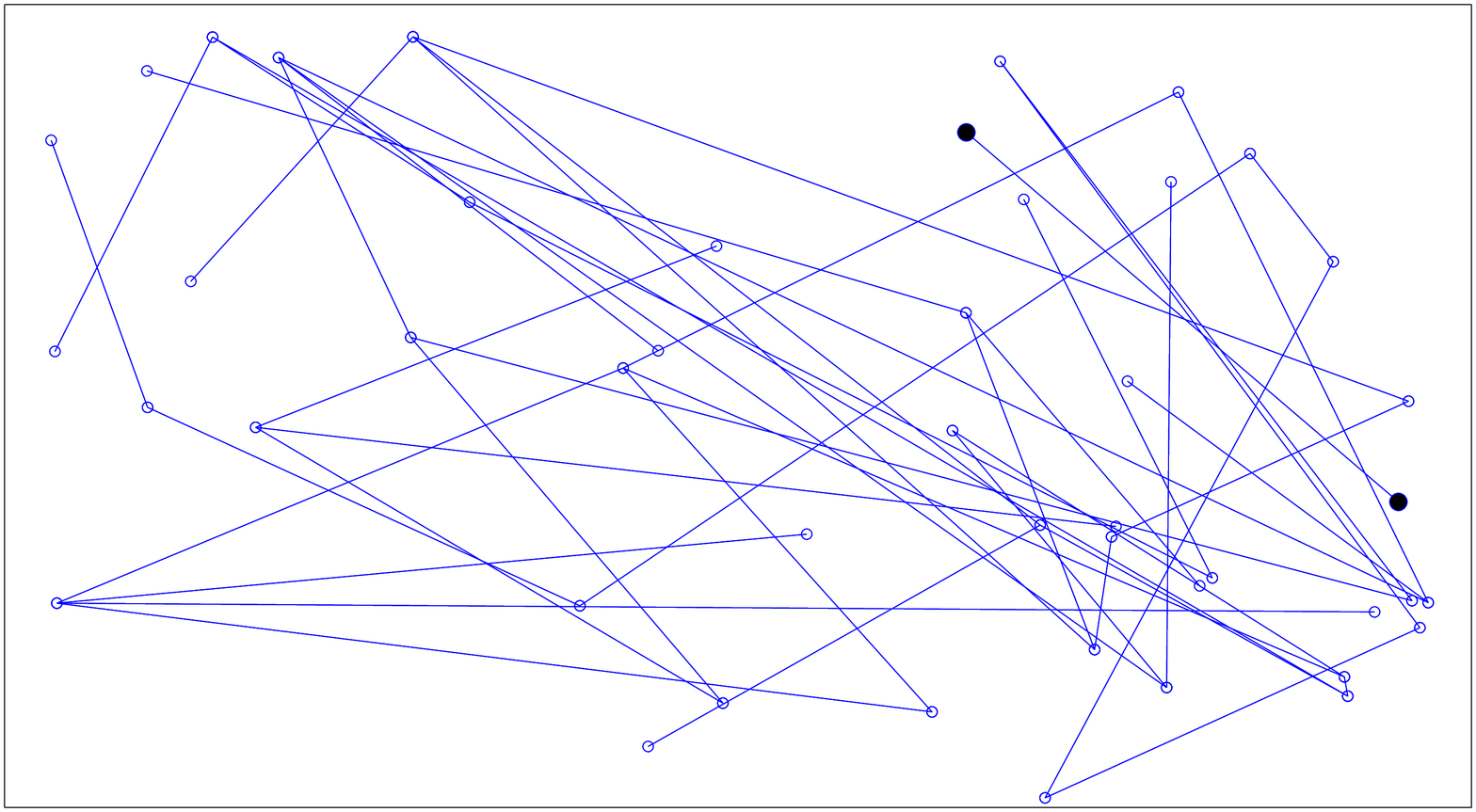} \caption{The intersection
$\mathbb{H \cap G}(n;\theta)$
         of the graphs in Figure \ref{figure:channel}
         and Figure \ref{figure:pairwise_graph} -- The graph
         is disconnected as the marked nodes form a component!
} \label{figure:intersection}
\end{figure}

When using random graph models to study networks, situations arise
where the notion of adjacency between nodes reflects multiple
constraints. This can be so even when dealing with networks other
than WSNs. As was the case here, such circumstances call for
studying models which are constructed by taking the intersection
of multiple random graphs. However, as pointed out earlier, the
availability of results for each component model does not
necessarily imply the availability of results for the intersection
of these models; see the examples provided in the previous
section.

Figures \ref{figure:channel}-\ref{figure:intersection} can help
better understand the relevant issues as to why this is so: Figure
\ref{figure:channel} provides a sample of an ER graph
$\mathbb{G}(n,p)$ with $n=200$ and $p=0.2$. As would be expected
from the classical results, the obtained graph is very densely
connected. Similarly, Figure \ref{figure:pairwise_graph} provides
a sample of the pairwise random graph $\mathbb{H}(n;K)$ with
$n=200$ and $K=5$. In line with Theorem
\ref{thm:OneLaw+ForConnectivity}, the obtained graph is connected.
On the other hand, the graph formed by intersecting these graphs
turn out to be {\em disconnected} as shown in Figure
\ref{figure:intersection}.

To drive this point further, consider the constant parameter case
for the models $\mathbb{H}(n;K)$ and $\mathbb{G}(n;p)$, a case
which cannot be recovered from either Theorem
\ref{thm:OneLaw+NodeIsolation} or Theorem
\ref{thm:OneLaw+Connectivity}. Nevertheless, Theorem
\ref{thm:OneLaw+ForConnectivity} yields
\[
\lim_{n \rightarrow \infty } \bP{\mathbb{H}(n;K)~\mbox{ is
connected} } = 1, \quad K \geq 2
\]
while it well known \cite{Bollobas} that
\[
\lim_{n \rightarrow \infty } \bP{\mathbb{G}(n;p)~\mbox{ is
connected} } = 1. \quad 0 < p < 1.
\]
However, it can be shown that
\begin{equation}
\lim_{n \rightarrow \infty } \bP{ \mathbb{H \cap
G}(n;\theta)~\mbox{contains} \mbox{~no~isolated~nodes} } = 0
\label{eq:ConstantAbsenceOfIsolatedNodes}
\end{equation}
whence
\begin{equation}
\lim_{n \rightarrow \infty } \bP{\mathbb{H \cap
G}(n;\theta)~\mbox{is connected} } = 0
\label{eq:ConstantConnectivity}
\end{equation}
for the same ranges of values for $p$ and $K$; for details see the
discussion at the end of Section
\ref{sec:ProofPropositionTechnical2}. This clearly provides a {\em
non-trivial} example (one that is not for an ER intersecting an ER
graph) where the intersection of two random graphs is indeed
a.a.s. {\em not} connected although each of the components is
a.a.s. connected.

\section{A proof of Theorem \ref{thm:OneLaw+NodeIsolation}}
\label{sec:ProofTheoremNodeIsolation}

We prove Theorem  \ref{thm:OneLaw+NodeIsolation} by the method of
first and second moments \cite[p.  55]{JansonLuczakRucinski}
applied to the total number of isolated nodes in $\mathbb{H \cap
G}(n;\theta)$. First some notation: Fix $n=2,3, \ldots $ and
consider $\theta = (K,p)$ with $p$ in $(0,1)$ and positive integer
$K$ such that $K < n$. With
\[
\chi_{n,i}(\theta) := \1{ {\rm Node~}i~{\rm is~isolated~in~}
                         \mathbb{H \cap G}(n;\theta) }
\]
for each $i=1, \ldots , n$, the number of isolated nodes in
$\mathbb{H \cap G}(n;\theta)$ is simply given by
\[
I_n (\theta) := \sum_{i=1}^n \chi_{n,i}(\theta).
\]
The random graph $\mathbb{H \cap G}(n;\theta)$ has no isolated
nodes if and only if $I_n (\theta) = 0$.

The method of first moment \cite[Eqn (3.10), p.
55]{JansonLuczakRucinski} relies on the well-known bound
\begin{equation}
1 - \bE{ I_n (\theta) } \leq \bP{  I_n (\theta) = 0 }
\label{eq:FirstMoment}
\end{equation}
while the method of second moment \cite[Remark 3.1, p.
55]{JansonLuczakRucinski} has its starting point in the inequality
\begin{equation}
\bP{  I_n (\theta) = 0 } \leq 1 - \frac{ \bE{ I_n (\theta)}^2}{
\bE{ I_n (\theta) ^2} }. \label{eq:SecondMoment}
\end{equation}

The rvs $\chi_{n,1}(\theta), \ldots , \chi_{n,n} (\theta)$ being
exchangeable, we find
\begin{equation}
\bE{ I_n (\theta)} = n \bE{ \chi_{n,1} (\theta) }
\label{eq:FirstMomentExpression}
\end{equation}
and
\begin{eqnarray}\label{eq:SecondMomentExpression}
\bE{ I_n (\theta)^2 } = n \bE{ \chi_{n,1} (\theta) }
 +  n(n-1) \bE{ \chi_{n,1}
(\theta)  \chi_{n,2} (\theta) } \nonumber
\end{eqnarray}
by the binary nature of the rvs involved. It then follows that
\begin{eqnarray}
\frac{ \bE{ I_n (\theta)^2 }}{ \bE{ I_n (\theta) }^2 } &=& \frac{
1}{ n\bE{ \chi_{n,1} (\theta) } }
\nonumber \\
& & + \frac{n-1}{n} \cdot \frac{\bE{ \chi_{n,1} (\theta)
\chi_{n,2} (\theta) }}
     {\left (  \bE{ \chi_{n,1} (\theta) } \right )^2 }.
\label{eq:SecondMomentRatio}
\end{eqnarray}

From (\ref{eq:FirstMoment}) and (\ref{eq:FirstMomentExpression})
we see that the one-law $\lim_{n\to \infty} \bP{I_n(\theta_n) = 0}
= 1$ will be established if we show that
\begin{equation}
\lim_{n \to \infty} n \bE{ \chi_{n,1} (\theta_n) }= 0.
\label{eq:OneLaw+NodeIsolation+convergence}
\end{equation}
It is also plain from (\ref{eq:SecondMoment}) and
(\ref{eq:SecondMomentExpression}) that the zero-law $\lim_{n \to
\infty} \bP{I_n(\theta_n) = 0} = 0$ holds if
\begin{equation}
\lim_{n \to \infty} n \bE{ \chi_{n,1} (\theta_n) }= \infty
\label{eq:OneLaw+NodeIsolation+convergence2}
\end{equation}
and
\begin{equation}
\limsup_{n \to \infty} \left( \frac{\bE{ \chi_{n,1} (\theta_n)
\chi_{n,2} (\theta_n) }}
     {\left (  \bE{ \chi_{n,1} (\theta_n) } \right )^2 }
\right) \leq 1. \label{eq:ZeroLaw+NodeIsolation+convergence}
\end{equation}

The proof of Theorem \ref{thm:OneLaw+NodeIsolation} passes through
the next two technical propositions which establish
(\ref{eq:OneLaw+NodeIsolation+convergence}),
(\ref{eq:OneLaw+NodeIsolation+convergence2}) and
(\ref{eq:ZeroLaw+NodeIsolation+convergence}) under the appropriate
conditions on the scaling $\theta: \mathbb{N}_0 \rightarrow
\mathbb{N}_0 \times (0,1)$.

\begin{proposition}
{\sl Consider scalings $K: \mathbb{N}_0 \rightarrow \mathbb{N}_0$
and $p: \mathbb{N}_0 \rightarrow (0,1)$ such that
(\ref{eq:scalinglaw}) holds for some $c>0$. Assume also that
$\lim_{n \to \infty}p_n=p^\star$ exists. Then, we have
\begin{equation}
\lim_{n \rightarrow \infty } n\bE{ \chi_{n,1} (\theta_n) } = \left
\{
\begin{array}{ll}
0 & \mbox{if~ $c > \tau(p^\star)$} \\
  &                      \\
\infty & \mbox{if~$c < \tau(p^\star)$}
\end{array}
\right . \label{eq:NodeIsolation+FirstMoment}
\end{equation}
where the threshold $\tau(p^\star)$ is given by
(\ref{eq:threshold}). } \label{prop:Technical1}
\end{proposition}
A proof of Proposition \ref{prop:Technical1} is given in Section
\ref{sec:ProofPropositionTechnical1}.

\begin{proposition}
{\sl Consider scalings $K: \mathbb{N}_0 \rightarrow \mathbb{N}_0$
and $p: \mathbb{N}_0 \rightarrow (0,1)$ such that
(\ref{eq:scalinglaw}) holds for some $c>0$. Assume also that
$\lim_{n \to \infty}p_n=p^\star$ exists. Then, we have
(\ref{eq:ZeroLaw+NodeIsolation+convergence}) whenever $p^ \star <
1$. } \label{prop:Technical2}
\end{proposition}

A proof of Proposition \ref{prop:Technical2} can be found in
Section \ref{sec:ProofPropositionTechnical2}. To complete the
proof of Theorem \ref{thm:OneLaw+NodeIsolation}, pick a scaling
$\theta: \mathbb{N}_0 \rightarrow \mathbb{N}_0 \times (0,1) $ such
that (\ref{eq:scalinglaw}) holds for some $c>0$ and $\lim_{n \to
\infty}p_n=p^\star$ exists. Under the condition $c> \tau(p^
\star)$ we get (\ref{eq:OneLaw+NodeIsolation+convergence}) from
Proposition \ref{prop:Technical1}, and the one-law $\lim_{n\to
\infty} \bP{I_n(\theta_n) = 0} = 1$ follows. Next, assume that $c<
\tau(p^ \star)$ -- This case is possible only if $p ^ \star < 1$
since $\tau(1)=0$ as seen at (\ref{eq:threshold}). When
$p^\star<1$, we obtain
(\ref{eq:OneLaw+NodeIsolation+convergence2}) and
(\ref{eq:ZeroLaw+NodeIsolation+convergence}) with the help of
Propositions \ref{prop:Technical1} and \ref{prop:Technical2},
respectively. The conclusion $\lim_{n \to \infty}
\bP{I_n(\theta_n) = 0} = 0$ is now immediate.

\section{A preparatory result}
\label{sec:PreparatoryResult}

Fix $n=2,3, \ldots $ and consider $\theta = (K,p)$ with $p$ in
$(0,1)$ and positive integer $K$ such that $K < n$. Under the
enforced assumptions, for all $i=1, \ldots , n$, we easily see
that
\begin{equation}
\bE{ \chi_{n,i} (\theta) } = \bE{(1-p)^{D_{n,i}}}
\label{eq:BeforeEvaluationFirstMoment}
\end{equation}
where $D_{n,i}$ denotes the degree of node $i$ in
$\mathbb{H}(n;K)$. Note that
\begin{eqnarray}
D_{n,i} &=& K + \sum_{j=1, j \notin \Gamma_{n,i} \cup\{i\}}^{n}
\1{i \in \Gamma_{n,j}}. \label{eq:degree_i}
\end{eqnarray}
By independence, since
\[
| \{ j=1, \ldots , n : \ j \notin \Gamma_{n,i} \cup\{i\} \} | = n-
K -1,
\]
the second term in (\ref{eq:degree_i}) is a binomial rv with
$n-K-1$ trials and success probability given by
\begin{equation}
\bP{i \in \Gamma_{n,j} } = \frac{{{n-2}\choose{K-1}}}{
{{n-1}\choose{K}}} =\frac{K}{n-1}, \label{eq:ProbabilityBasic1}
\end{equation}
whence
\begin{equation}
\bE{ \chi_{n,i} (\theta) } = \left(1-p\right)^K \cdot \left(1 -
\frac{pK}{n-1}\right)^{n-K-1} . \label{eq:EvaluationFirstMoment}
\end{equation}

The proof of Proposition \ref{prop:Technical1} uses a somewhat
simpler form of the expression (\ref{eq:EvaluationFirstMoment})
which we develop next.

\begin{lemma}
{\sl Consider scalings $K: \mathbb{N}_0 \rightarrow \mathbb{N}_0$
and $p: \mathbb{N}_0 \rightarrow (0,1)$ such that
(\ref{eq:scalinglaw}) holds for some $c>0$. It holds that
\begin{equation}
n \bE{ \chi_{n,1} (\theta_n) } = e^{\alpha_n + o(1) }
\label{eq:PreparatoryLemma} \quad n=1,2, \ldots
\end{equation}
with
\begin{equation}
\alpha_n := (1-c_n) \log n  + K_n (p_n + \log (1-p_n))
\label{eq:alpha_n}
\end{equation}
where the sequence $c: \mathbb{N}_0 \rightarrow \mathbb{R}$ is the
one appearing in the form (\ref{eq:scalinglawEquivalent}) of the
condition (\ref{eq:scalinglaw}). } \label{lem:PreparatoryLemma}
\end{lemma}

In what follows we make use of the decomposition
\begin{equation}
\log ( 1 - x ) = -x - \Psi (x), \quad 0 \leq x < 1
\label{eq:LogDecomposition}
\end{equation}
with
\[
\Psi(x) := \int_0^x \frac{t}{1-t} dt
\]
on that range. Note that
\[
\lim_{x \downarrow 0} \frac{ \Psi(x) }{x^2} = \frac{1}{2}.
\]

\myproof Consider a scaling $\theta: \mathbb{N}_0 \rightarrow
\mathbb{N}_0 \times (0,1)$ such that (\ref{eq:scalinglaw}) holds
for some $c>0$ and assume the existence of the limit $\lim_{n \to
\infty}p_n=p^\star$. Replacing $\theta$ by $\theta_n$ in
(\ref{eq:EvaluationFirstMoment}) for each $n=2,3, \ldots $ we get
\begin{equation}
n \bE{ \chi_{n,1} (\theta_n) } = e^{\beta_n}
\end{equation}
with $\beta_n$ given by
\[
\beta_n = \log n + K_n \log (1-p_n) - \gamma_n
\]
with
\[
\gamma_n := -(n-K_n-1) \log \left ( 1 - \frac{p_n K_n}{n-1} \right
) .
\]
The decomposition (\ref{eq:LogDecomposition}) now yields
\begin{eqnarray}
\gamma_n &:=& (n-K_n-1) \left ( \frac{p_nK_n}{n-1} + \Psi \left (
\frac{p_nK_n}{n-1} \right ) \right )
\nonumber \\
&=& \left ( 1 - \frac{K_n}{n-1} \right ) K_np_n + (n-K_n-1) \Psi
\left ( \frac{p_nK_n}{n-1} \right )
\nonumber \\
&=& - K_np_n + \left ( 2 - \frac{K_n}{n-1} \right ) K_np_n
\nonumber \\
& & \quad \quad + (n-K_n-1) \Psi \left ( \frac{p_nK_n}{n-1} \right
)
\nonumber \\
&=& - K_np_n + c_n \log n + (n-K_n-1) \Psi \left (
\frac{p_nK_n}{n-1} \right ) \nonumber
\end{eqnarray}
where the last step used the form (\ref{eq:scalinglawEquivalent})
of the condition (\ref{eq:scalinglaw}) on the scaling. Reporting
this calculation into the expression for $\beta_n$ we find
\[
\beta_n = \alpha_n - (n-K_n-1) \Psi \left ( \frac{p_nK_n}{n-1}
\right ).
\]

Lemma \ref{lem:PreparatoryLemma} will be established if we show
that
\begin{equation}
\lim_{ n \rightarrow \infty} (n-K_n-1) \Psi \left (
\frac{p_nK_n}{n-1} \right ) = 0 .
\label{eq:PreparatoryLemmaKeyLimit}
\end{equation}
To that end, for each $n=2,3, \ldots $ we note that
\[
p_n K_n \leq p_n\left(2K_n-\frac{K_n ^ 2}{n-1}\right) \leq 2 p_n
K_n
\]
since $K_n < n$. The condition (\ref{eq:scalinglawEquivalent})
implies
\begin{equation}
\frac{c_n}{2}  \log n \leq p_n K_n \leq c_n \log n,
\label{eq:usefulinequality}
\end{equation}
and it is now plain that
\[
\lim_{n \to \infty} \frac{p_n K_n}{n-1} = 0 \quad \mbox{and} \quad
\lim_{n \to \infty} (n-K_n-1) \frac{p_n ^ 2 K_n ^ 2}{(n-1)^2} = 0.
\]
Invoking the behavior of $\Psi(x)$ at $x=0$ mentioned earlier, we
conclude from these facts that
\begin{equation}
\lim_{n \to \infty} \left ( (n-K_n-1)\frac{p_n ^ 2 K_n ^
2}{(n-1)^2} \right ) \left( \frac{\Psi\left(\frac{p_n
K_n}{n-1}\right) }
     {\left(\frac{p_n K_n}{n-1}\right)^2}\right)
= 0 . \label{eq:UsefulLimit}
\end{equation}
This establishes (\ref{eq:PreparatoryLemmaKeyLimit}) and the proof
of Lemma \ref{lem:PreparatoryLemma} is completed. \myendpf

\section{A proof of Proposition \ref{prop:Technical1}}
\label{sec:ProofPropositionTechnical1}

In view of Lemma \ref{lem:PreparatoryLemma}, Proposition
\ref{prop:Technical1} will be established if we show
\begin{equation}
\lim_{n \rightarrow \infty } \alpha_n = \left \{
\begin{array}{ll}
-\infty & \mbox{if~ $c > \tau(p^\star)$} \\
  &                      \\
+\infty & \mbox{if~$c < \tau(p^\star)$.}
\end{array}
\right . \label{eq:alpha_N_establish}
\end{equation}

To see this, first note from (\ref{eq:LogDecomposition}) that for
each $n = 1,2, \ldots $, we have $p_n + \log (1-p_n) \leq 0$ and
the lower bound in (\ref{eq:usefulinequality}) implies
\begin{eqnarray}
\alpha_n &\leq& (1-c_n) \log n  + c_n \left ( \frac{\log n}{2 p_n}
\right ) \cdot \left ( p_n + \log (1-p_n) \right )
\nonumber \\
&=& \left ( 1 - \frac{c_n}{2} \left( 1-\frac{\log(1-p_n)}{p_n}
\right) \right) \cdot \log n .
\end{eqnarray}
Letting $n$ go to infinity in this last expression, we get
$\lim_{n \to \infty} \alpha_n=-\infty$ whenever
\begin{equation}
c > \lim_{n \to \infty} \frac{2}{1-\frac{\log(1-p_n)}{p_n}} =
\tau(p^\star)
\end{equation}
since $\lim_{n \rightarrow \infty} c_n = c$.

Next, we show that if $c < \tau (p^ \star)$, then $\lim_{n \to
\infty} \alpha_n=+\infty$. We only need to consider the case $0
\leq p^ \star < 1$ since $\tau(1)=0$ and the constraint $c < \tau
(1) $ is vacuous. We begin by assuming $p^ \star = 0$, in which
case for each $n=2,3, \ldots$, we have
\begin{eqnarray}
\alpha_n &=& (1-c_n) \log n  + K_n (p_n + (- p_n - \Psi(p_n)))
\nonumber \\
&=& (1-c_n) \log n- K_n \Psi(p_n)
\nonumber \\
&=& (1-c_n) \log n- \left(\frac{\Psi(p_n)}{p_n ^2}\right) \cdot
K_n p_n^2
\nonumber \\
&\geq& (1-c_n) \log n- c_n \log n \cdot \left(\frac{\Psi(p_n)}{p_n
^2}\right) p_n
\nonumber \\
&=& \log n \cdot \left( 1 - c_n \left ( 1+
\left(\frac{\Psi(p_n)}{p_n ^2}\right) p_n \right ) \right)
\end{eqnarray}
with the inequality following from the upper bound in
(\ref{eq:usefulinequality}). Let $n$ grow large in the last
expression. Since we have assumed $\lim_{n \to \infty} p_n = 0$,
we get
\[
\lim_{n \to \infty}p_n \left(\frac{\Psi(p_n)}{p_n ^2}\right) =0,
\]
and the desired conclusion $\lim_{n \to \infty} \alpha_n=+\infty$
is obtained whenever $c<1=\tau(0)$ upon using $\lim_{n \rightarrow
\infty} c_n = c$.

Finally we assume $0<p^\star<1$. For each $\varepsilon >0$, there
exists a finite positive integer $n^\star (\varepsilon)$ such that
$p_n \geq (1-\varepsilon)p^ \star$ when $n\geq n^ \star
(\varepsilon)$. On that range the upper bound in
(\ref{eq:usefulinequality}) yields
\[
K_n \leq \frac{c}{(1-\varepsilon)p^\star} \cdot \log n,
\]
whence the conclusions $K_n^2= o(n)$ and
\[
p_n \left(2K_n -\frac{K_n^2}{n-1}\right) = 2K_np_n + o(1)
\]
follow. Comparing this last fact against the lefthand side of
(\ref{eq:scalinglawEquivalent}) yields
\[
K_np_n = \frac{c_n}{2} \log n + o(1),
\]
so that
\begin{equation}
K_np_n \sim \frac{c_n}{2} \log n . \label{eq:Equivalence22}
\end{equation}

From (\ref{eq:alpha_n}) it follows that
\[
\frac{\alpha_n}{\log n } = (1-c_n) + \left ( 1 + \frac{ \log (1 -
p_n) } { p_n } \right ) \cdot \frac{K_n p_n}{\log n}
\]
for all $n$ sufficiently large. Letting $n$ go to infinity in this
last expression and using (\ref{eq:Equivalence22}) with the
earlier remarks, we readily conclude
\[
\lim_{n \rightarrow \infty} \frac{\alpha_n}{\log n } = (1-c) +
\frac{c}{2} \left ( 1 + \frac{ \log (1 - p^\star) } { p^\star }
\right ) = 1 - \frac{c}{\tau(p^\star)}
\]
where the last step follows by direct inspection. It is now clear
that $\lim_{n \rightarrow \infty} \alpha_n = \infty$ when $c <
\tau(p^\star)$ with $0 < p^\star < 1$. The proof of Proposition
\ref{prop:Technical1} is now completed. \myendpf

\section{Negative dependence and consequences}
\label{sec:NegativeDependence}

Fix positive integers $n=2,3, \ldots $ and $K$ with $K < n$.
Several properties of the $\{0,1\}$-valued rvs
\begin{equation}
\left \{ \1 { j \in \Gamma_{n,i} }, \quad
\begin{array}{c}
i \neq j \\
i,j =1, \ldots , n
\end{array}
\right \} \label{eq:NegativeAssociation1}
\end{equation}
and
\begin{equation}
\left \{ \1 { j \in \Gamma_{n,i} \ \vee \ i \in \Gamma_{n,j} },
\quad
\begin{array}{c}
i \neq j \\
i,j =1, \ldots , n
\end{array}
\label{eq:NegativeAssociation3} \right \}
\end{equation}
will play a key role in some of the forthcoming arguments.

\subsection{Negative association}
\label{subsec:Definitions}

The properties of interest can be couched in terms of {\em
negative association}, a form of negative correlation introduced
to Joag-Dev and Proschan \cite{Joag-DevProschan}. We first develop
the needed definitions and properties: Let $\{ X_\lambda , \
\lambda \in \Lambda \}$ be a collection of $\mathbb{R}$-valued rvs
indexed by the finite set $\Lambda$. For any non-empty subset $A$
of $\Lambda$, we write $X_A$ to denote the
$\mathbb{R}^{|A|}$-valued $X_A = ( X_\lambda , \ \lambda \in A )$.
The rvs $\{ X_\lambda , \ \lambda \in \Lambda \}$ are then said to
be {\sl negatively associated} if for any non-overlapping subsets
$A$ and $B$ of $\Lambda$ and for any monotone increasing mappings
$\varphi : \mathbb{R}^{|A|} \rightarrow \mathbb{R}$ and $\psi :
\mathbb{R}^{|B|} \rightarrow \mathbb{R}$, the covariance
inequality
\begin{equation}
\bE{ \varphi( X_A ) \psi( X_B ) } \leq \bE{ \varphi( X_A )} \bE{
\psi( X_B ) } \label{eq:NegativeAssociationDefn}
\end{equation}
holds whenever the expectations in
(\ref{eq:NegativeAssociationDefn}) are well defined and finite.
Note that $\varphi$ and $\psi$ need only be monotone increasing on
the support of $X_A$ and $X_B$, respectively.

This definition has some easy consequences to be used repeatedly
in what follows: The negative association of $\{ X_\lambda , \
\lambda \in \Lambda \}$ implies the negative association of the
collection $\{ X_\lambda , \ \lambda \in \Lambda^\prime \}$ where
$\Lambda^\prime$ is any subset of $\Lambda$. It is also well known
\cite[P2, p. 288]{Joag-DevProschan} that the negative association
of the rvs $\{ X_\lambda , \ \lambda \in \Lambda \}$ implies the
inequality
\begin{equation}
\bE{ \prod_{\lambda \in A} f_\lambda (X_\lambda ) } \leq
\prod_{\lambda \in A} \bE{ f_\lambda (X_\lambda ) }
\label{eq:NegativeAssociationConsequence}
\end{equation}
where $A$ is a subset of $\Lambda$ and the collection $\{
f_\lambda , \ \lambda \in A \}$ of mappings $\mathbb{R}
\rightarrow \mathbb{R}_+$ are all monotone increasing; by
non-negativity all the expectations exist and finiteness is moot.

We can apply these ideas to collections of indicator rvs, namely
for each $\lambda$ in $\Lambda$, $X_\lambda = \1{ E_\lambda }$ for
some event $E_\lambda$. From the definitions, it is easy to see
that if the rvs $\{ \1{E_\lambda } , \ \lambda \in \Lambda \}$ are
negatively associated, so are the rvs $\{ \1{E^c_\lambda } , \
\lambda \in \Lambda \}$. Moreover, for any subset $A$ of
$\Lambda$, we have
\begin{equation}
\bP{ E_\lambda , \ \lambda \in A } \leq \prod_{\lambda \in A} \bP{
E_\lambda }. \label{eq:KeyConsequenceForProbabilities}
\end{equation}
This follows from (\ref{eq:NegativeAssociationConsequence}) by
taking $f_\lambda (x) = x^+$ on $\mathbb{R}$ for each $\lambda$ in
$\Lambda$.

\subsection{Useful consequences}
\label{subsec:UsefulConsequences}

A key observation for our purpose is as follows: For each $i=1,
\ldots , n$, the rvs
\begin{equation}
\{ \1 { j \in \Gamma_{n,i} }, \ j \in {\cal N}_{-i} \}
\label{eq:NegativeAssociation2}
\end{equation}
form a collection of negatively associated rvs. This is a
consequence of the fact that the random set $\Gamma_{n,i}$
represents a random sample (without replacement) of size $K$ from
${\cal N}_{-i}$; see \cite[Example 3.2(c)]{Joag-DevProschan} for
details.

The $n$ collections (\ref{eq:NegativeAssociation2}) are mutually
independent, so that by the \lq \lq closure under products"
property of negative association \cite[P7, p.
288]{Joag-DevProschan} \cite[p. 35]{DubhashiPanconesi}, the rvs
(\ref{eq:NegativeAssociation1}) also form a collection of
negatively associated rvs.

Hence, by taking complements, the rvs
\begin{equation}
\left \{ \1 { j \notin \Gamma_{n,i} }, \quad
\begin{array}{c}
i \neq j \\
i,j =1, \ldots , n
\end{array}
\label{eq:NegativeAssociation3Complement} \right \}
\end{equation}
also form a collection of negatively associated rvs. With distinct
$i,j=1, \ldots , n$, we note that
\begin{eqnarray}
\1{ i \notin \Gamma_{n,j}, j \notin \Gamma_{n,i} } = f \left ( \1{
i \notin \Gamma_{n,j} }, \1{ j \notin \Gamma_{n,i} } \right )
\end{eqnarray}
with mapping $f: \mathbb{R}^2 \rightarrow \mathbb{R}$ given by
$f(x,y) = x^+y^+$ for all $x,y$ in $\mathbb{R}$. This mapping
being non-decreasing on $\mathbb{R}^2$, it follows  \cite[P6, p.
288]{Joag-DevProschan} that the rvs
\begin{equation}
\left \{ \1 { j \notin \Gamma_{n,i}, i \notin \Gamma_{n,j} },
\quad
\begin{array}{c}
i \neq j \\
i,j =1, \ldots , n
\end{array}
\label{eq:NegativeAssociation5} \right \}
\end{equation}
are also negatively associated. Taking complements one more time,
we see that the rvs (\ref{eq:NegativeAssociation3}) are also
negatively associated.

For each $k=1,2$ and $j=3, \ldots , n$, we shall find it useful to
define
\[
u_{n,j,k}(\theta) := \bE{ (1-p)^{ \1{ k \in \Gamma_{n,j} } } }
\]
and
\[
b_{n,j} (\theta) := \bE{ (1-p)^ { \1{ 1 \in \Gamma_{n,j} } + \1{ 2
\in \Gamma_{n,j} } } }.
\]
Under the enforced assumptions, we have $b_{n,3}(\theta) = \ldots
= b_{n,n}(\theta) \equiv b_n(\theta) $ and $u_{n,3,1}(\theta) =
\ldots = u_{n,n,1}(\theta) = u_{n,3,2}(\theta) = \ldots =
u_{n,n,2}(\theta) \equiv u_n (\theta)$.

Before computing either one of the quantities $u_n(\theta)$ and
$b_n(\theta)$, we note that
\begin{equation}
b_n(\theta) \leq u_n(\theta)^2 . \label{eq:BleqU^2}
\end{equation}
This is a straightforward consequence of the negative association
of the rvs (\ref{eq:NegativeAssociation1}) -- In
(\ref{eq:NegativeAssociationDefn}), with $A$ and $B$ singletons,
use the increasing functions $\varphi, \psi: \mathbb{R}
\rightarrow \mathbb{R} : x \rightarrow -(1-p)^x$.

Using (\ref{eq:ProbabilityBasic1}) we get
\begin{eqnarray}
u_n(\theta) &=& (1-p) \frac{K}{n-1} + \left ( 1 - \frac{K}{n-1}
\right )
\nonumber \\
&=& 1 - p \frac{K}{n-1} . \label{eq:Compute1}
\end{eqnarray}
An expression for $b_n(\theta)$ is available but will not be
needed due to the availability of (\ref{eq:BleqU^2}).

\section{A proof of
         Proposition \ref{prop:Technical2}}
\label{sec:ProofPropositionTechnical2}

As expected, the first step in proving Proposition
\ref{prop:Technical2} consists in evaluating the cross moment
appearing in the numerator of
(\ref{eq:ZeroLaw+NodeIsolation+convergence}). Fix $n=2,3, \ldots $
and consider $\theta = (K,p)$ with $p$ in $(0,1)$ and positive
integer $K$ such that $K < n$. Define the $\mathbb{N}_0$-valued
rvs $B_n(\theta)$ and $U_n(\theta)$ by
\begin{equation}
B_n(\theta) := \sum_{j=3}^n \1{ j \not \in \Gamma_{n,1} } \1{ j
\not \in \Gamma_{n,2} } \label{eq:B_n}
\end{equation}
and
\begin{eqnarray}
U_n(\theta) &:=& \sum_{j=3}^n \1{ j \not \in \Gamma_{n,1} } \1{ j
\in \Gamma_{n,2} } \label{eq:U_n}
\\
& & + \sum_{j=3}^n \1{ j \not \in \Gamma_{n,2} } \1{ j \in
\Gamma_{n,1} } . \nonumber
\end{eqnarray}

\begin{proposition}
{\sl Fix $n=2,3, \ldots $. For any $p$ in $(0,1)$ and positive
integer $K$ such that $K < n$, we have
\begin{eqnarray}
\lefteqn{\bE{ \chi_{n,1} (\theta) \chi_{n,2} (\theta) }} &&
\label{eq:EvalCrossMoment} \\
&=& (1-p)^{2K} \bE{ \frac{ b_n(\theta)^{B_n(\theta)} \cdot
u_n(\theta)^{U_n(\theta)} }
     { (1-p)^{ \1{ 2 \in \Gamma_{n,1} , 1 \in \Gamma_{n,2} } } }
} \nonumber
\end{eqnarray}
where the rvs $B_n(\theta)$ and $U_n(\theta)$ given by
(\ref{eq:B_n}) and (\ref{eq:U_n}), respectively. }
\label{prop:EvalCrossMoment}
\end{proposition}

A proof of Proposition \ref{prop:EvalCrossMoment} is available in
Appendix \ref{App:A}. Still in the setting of Proposition
\ref{prop:EvalCrossMoment}, we can use (\ref{eq:BleqU^2}) in
conjunction with (\ref{eq:EvalCrossMoment}) to get
\begin{eqnarray}
\lefteqn{ \bE{ \chi_{n,1} (\theta) \chi_{n,2} (\theta) } } & &
\label{eq:EvalCrossMomentBound} \\
&\leq& (1-p)^{2K} \bE{ \frac{ u_n(\theta)^{2B_n(\theta) +
U_n(\theta)} }
     { (1-p)^{ \1{ 2 \in \Gamma_{n,1} , 1 \in \Gamma_{n,2} } } }
}. \nonumber
\end{eqnarray}
It is plain that
\begin{eqnarray}
\lefteqn{ 2B_n(\theta) + U_n(\theta) } & &
\nonumber \\
&=& \sum_{j=3}^n \1{ j \not \in \Gamma_{n,1} } + \sum_{j=3}^n \1{
j \not \in \Gamma_{n,2} } . \nonumber
\end{eqnarray}
We note that
\begin{eqnarray}
\sum_{j=3}^n \1{ j \not \in \Gamma_{n,1} } &=& \sum_{j=2}^n \1{ j
\not \in \Gamma_{n,1} } - \1{ 2 \not \in \Gamma_{n,1} }
\nonumber \\
&=& (n-1-K) - \left ( 1 - \1{ 2 \in \Gamma_{n,1} } \right )
\nonumber \\
&=& (n-2-K)  + \1{ 2 \in \Gamma_{n,1} } \nonumber
\end{eqnarray}
and
\begin{eqnarray}
\sum_{j=3}^n \1{ j \not \in \Gamma_{n,2} } = (n-2-K)  + \1{ 1 \in
\Gamma_{n,2} } \nonumber
\end{eqnarray}
by similar arguments. The expression
\begin{eqnarray}
\lefteqn{ 2B_n(\theta) + U_n(\theta) } & &
\\
& =& 2 (n-2-K) + \1{ 2 \in \Gamma_{n,1} } + \1{ 1 \in \Gamma_{n,2}
} \nonumber
\end{eqnarray}
now follows, and we find
\begin{eqnarray}
\lefteqn{ \bE{ \chi_{n,1} (\theta) \chi_{n,2} (\theta) } } & &
\label{eq:EvalCrossMomentBound2} \\
&\leq& (1-p)^{2K} u_n(\theta)^{2 (n-2-K) } \cdot R_n(\theta)
\nonumber
\end{eqnarray}
with
\[
R_n (\theta) := \bE{  \frac{u_n(\theta) ^ {\1{ 2 \in \Gamma_{n,1}
} + \1{ 1 \in \Gamma_{n,2} } }} {(1-p)^{ \1{ 2 \in \Gamma_{n,1} ,
1 \in \Gamma_{n,2} } } } } .
\]

Next, with the help of (\ref{eq:EvaluationFirstMoment}) and
(\ref{eq:Compute1}) we conclude that
\begin{eqnarray}
\lefteqn{ \frac{\bE{ \chi_{n,1} ( \theta ) \chi_{n,2} ( \theta ) }
}
     {\left (  \bE{ \chi_{n,1} ( \theta ) } \right ) ^ 2 }
} & &
\nonumber \\
&\leq& \frac{ (1-p)^{2K} \cdot u_n(\theta)^{2 (n-2-K) }}
     { \left ( (1-p)^K \cdot u_n(\theta)^{n-1-K} \right )^2 }
\cdot R_n(\theta)
\nonumber \\
&=& u_n(\theta)^{-2} R_n(\theta)
\nonumber \\
&=& \bE{ \frac{u_n(\theta) ^ {\1{ 2 \in \Gamma_{n,1} }
                     + \1{ 1 \in \Gamma_{n,2} } - 2 }}
     {(1-p)^{ \1{ 2 \in \Gamma_{n,1} , 1 \in \Gamma_{n,2} } } }
}. \label{eq:EvaluationFirstMomentBound3}
\end{eqnarray}
Direct inspection readily yields
\begin{eqnarray}
\lefteqn{ \frac{u_n(\theta) ^ {\1{ 2 \in \Gamma_{n,1} }
                     + \1{ 1 \in \Gamma_{n,2} } - 2 }}
     {(1-p)^{ \1{ 2 \in \Gamma_{n,1} , 1 \in \Gamma_{n,2} } } }} &&
\\ \nonumber
&=& \left \{
\begin{array}{lll}
\frac{1}{1-p} & \mbox{if~ $2 \in \Gamma_{n,1} , 1 \in
\Gamma_{n,2}$} \\
  &                      \\
\left(1-\frac{p K}{n-1}\right)^{-2} & \mbox{if~$2 \not \in
\Gamma_{n,1} , 1 \not \in \Gamma_{n,2}$} \\
 &                      \\
\left(1-\frac{p K}{n-1}\right)^{-1} & \mbox{otherwise.}
\end{array}
\right .
\end{eqnarray}
Taking expectation and reporting into
(\ref{eq:EvaluationFirstMomentBound3}) we then find
\begin{eqnarray}
\lefteqn{ \frac{\bE{ \chi_{n,1} ( \theta ) \chi_{n,2} ( \theta ) }
}
     {\left (  \bE{ \chi_{n,1} ( \theta ) } \right ) ^ 2 }
} &&
\\ \nonumber
&\leq& \frac{1}{1-p} \bP{2 \in \Gamma_{n,1} , 1 \in \Gamma_{n,2}}
+ \left ( 1 - p \frac{K}{n-1} \right )^{-2}
\nonumber \\
&=& \frac{1}{1-p} \left ( \frac{K}{n-1} \right )^2 + \left ( 1 - p
\frac{K}{n-1} \right )^{-2} \label{eq:EvaluationFirstMomentBound6}
\end{eqnarray}
by a crude bounding argument.

Now consider a scaling $\theta: \mathbb{N}_0 \rightarrow
\mathbb{N}_0 \times (0,1)$ such that (\ref{eq:scalinglaw}) holds
for some $c>0$ and $\lim_{n \to \infty}p_n=p^\star < 1$. Replace
$\theta$ by $\theta_n$ in the bound
(\ref{eq:EvaluationFirstMomentBound6}) with respect to this
scaling. It is immediate that
(\ref{eq:ZeroLaw+NodeIsolation+convergence}) will be established
if we show that
\[
\lim_{n \rightarrow \infty}
\frac{1}{1-p_n}\left(\frac{K_n}{n-1}\right)^2 = 0
\]
and that
\[
\lim_{n \rightarrow \infty} \left ( 1 - p_n \frac{K_n}{n-1}
\right) = 1.
\]
These limits are an easy consequence of the inequalities
(\ref{eq:usefulinequality}) by virtue of the fact that $\lim_{n
\to \infty}p_n=p^\star < 1$. \myendpf

We close with a proof of
(\ref{eq:ConstantAbsenceOfIsolatedNodes}): Consider $\theta =
(K,p)$ with $p$ in $(0,1)$ and positive integer $K$. It follows
from (\ref{eq:EvaluationFirstMoment}) that
\[
\lim_{n \rightarrow \infty} \bE{ \chi_{n,1} (\theta) } =
\left(1-p\right)^K e^{-pK},
\]
whence $ \lim_{n \rightarrow \infty} \bE{ I_n(\theta) } = \infty$.
It also immediate from (\ref{eq:EvaluationFirstMomentBound6}) that
\[
\limsup_{n \rightarrow \infty} \frac{\bE{ \chi_{n,1} ( \theta )
\chi_{n,2} ( \theta ) } }
     {\left (  \bE{ \chi_{n,1} ( \theta ) } \right ) ^ 2 }
\leq 1 .
\]
The arguments outlined in Section
\ref{sec:ProofTheoremNodeIsolation} now yield
\[
\lim_{n \rightarrow \infty} \bP{ I_n(\theta) = 0 } = 0,
\]
and this establishes (\ref{eq:ConstantAbsenceOfIsolatedNodes}).
The conclusion (\ref{eq:ConstantConnectivity}) immediately
follows; see discussion at
(\ref{eq:FromConnectivityToNodeIsolation1}).

\section{A proof of Theorem \ref{thm:OneLaw+Connectivity} (Part I)}
\label{sec:ProofConnectivityI}

Fix $n=2,3, \ldots $ and consider $\theta = (K,p)$ with $p$ in
$(0,1)$ and positive integer $K$ such that $K < n$. We define the
events
\[
C_n(\theta) := \left [ \mathbb{H \cap G}(n; \theta)
\mbox{~is~connected} \right ]
\]
and
\[
I ( n ; \theta) := \left [ \mathbb{H \cap G}(n;\theta)
\mbox{~contains~no~isolated~nodes} \right ].
\]
If the random graph $\mathbb{H \cap G}(n; \theta)$ is connected,
then it does not contain any isolated node, whence $C_n(\theta)$
is a subset of $I(n;\theta)$, and the conclusions
\begin{equation}
\bP{ C_n(\theta) } \leq \bP{ I(n;\theta) }
\label{eq:FromConnectivityToNodeIsolation1}
\end{equation}
and
\begin{equation}
\bP{ C_n(\theta)^c } = \bP{ C_n(\theta)^c \cap I (n;\theta) } +
\bP{ I(n;\theta)^c } \label{eq:FromConnectivityToNodeIsolation2}
\end{equation}
obtain.

Taken together with Theorem \ref{thm:OneLaw+NodeIsolation}, the
relations (\ref{eq:FromConnectivityToNodeIsolation1}) and
(\ref{eq:FromConnectivityToNodeIsolation2}) pave the way to
proving Theorem \ref{thm:OneLaw+Connectivity}. Indeed, pick a
scaling $\theta: \mathbb{N}_0 \rightarrow \mathbb{N}_0 \times
(0,1)$ such that (\ref{eq:scalinglaw}) holds for some $c>0$ and
$\lim_{n \to \infty}p_n=p^\star$ exists. If $c < \tau(p^\star)$,
then $\lim_{n \rightarrow \infty} \bP{ I(n;\theta_n) } = 0$ by the
zero-law for the absence of isolated nodes, whence $\lim_{n
\rightarrow \infty} \bP{ C_n(\theta_n) } = 0$ with the help of
(\ref{eq:FromConnectivityToNodeIsolation1}). If $c>
\tau(p^\star)$, then $\lim_{n \rightarrow \infty} \bP{
I(n;\theta_n) } = 1$ by the one-law for the absence of isolated
nodes, and the desired conclusion $\lim_{n \rightarrow \infty }
\bP{ C_n(\theta_n) } = 1$ (or equivalently, $\lim_{n \rightarrow
\infty } \bP{ C_n(\theta_n)^c } = 0$) will follow via
(\ref{eq:FromConnectivityToNodeIsolation2}) if we show the
following:

\begin{proposition}
{\sl For any scaling $\theta: \mathbb{N}_0 \rightarrow
\mathbb{N}_0 \times (0,1) $ such that $\lim_{n \to
\infty}p_n=p^\star$ exists and (\ref{eq:scalinglaw}) holds for
some $c>\tau (p ^ \star)$, we have
\begin{equation}
\lim_{n \rightarrow \infty} \bP{ C_n(\theta_n)^c \cap I
(n;\theta_n) } = 0 . \label{eq:OneLawAfterReduction}
\end{equation}
} \label{prop:OneLawAfterReduction}
\end{proposition}

The proof of Proposition \ref{prop:OneLawAfterReduction} starts
below and runs through two more sections, namely Sections
\ref{sec:BoundingProbabilities} and \ref{sec:ProofConnectivityII}.
The basic idea is to find a sufficiently tight upper bound on the
probability in (\ref{eq:OneLawAfterReduction}) and then to show
that this bound goes to zero as $n$ becomes large. This approach
is similar to the one used for proving the one-law for
connectivity in ER graphs \cite[p. 164]{Bollobas}.

We begin by finding the needed upper bound: Fix $n=2,3, \ldots $
and consider $\theta = (K,p)$ with $p$ in $(0,1)$ and positive
integer $K$ such that $K < n$. For any non-empty subset $S$ of
nodes, i.e., $S \subseteq \{1, \ldots , n \}$, we define the graph
$\mathbb{H \cap G} (n;\theta) (S)$ (with vertex set $S$) as the
subgraph of $\mathbb{H \cap G} (n;\theta)$ restricted to the nodes
in $S$. We also say that $S$ is {\em isolated} in $\mathbb{H \cap
G} (n;\theta)$ if there are no edges (in $\mathbb{H \cap G}
(n;\theta)$) between the nodes in $S$ and the nodes in the
complement $S^c = \{ 1, \ldots , n \} - S$. This is characterized
by
\[
\Sigma_{n,i} \cap \Sigma_{n,j}
 = \emptyset \:\: \vee \:\: B_{ij}(p) = 0 ,
\quad i \in S , \ j \in S^c  .
\]

With each non-empty subset $S$ of nodes, we associate several
events of interest: Let $C_n (\theta ; S)$ denote the event that
the subgraph $\mathbb{H \cap G} (n;\theta) (S)$ is itself
connected. The event $C_n (\theta ; S)$ is completely determined
by the rvs $\{ K_i(\theta), \ i \in S \}$. We also introduce the
event $B_n (\theta ; S)$ to capture the fact that $S$ is isolated
in $\mathbb{H \cap G} (n;\theta)$, i.e.,
\begin{eqnarray}
\lefteqn{ B_n (\theta ; S) } & &
\nonumber \\
&:= & \left [ \Sigma_{n,i} \cap \Sigma_{n,j} = \emptyset \:\: \vee
\:\: B_{ij}(p) = 0 , \quad i \in S , \ j \in S^c \right ] .
\nonumber
\end{eqnarray}
Finally, we set
\[
A_n (\theta ; S) := C_n (\theta ; S) \cap B_n (\theta ; S) .
\]

The starting point of the discussion is the following basic
observation: If $\mathbb{H \cap G} (n;\theta)$ is {\em not}
connected and yet has {\em no} isolated nodes, then there must
exist a subset $S$ of nodes with $|S| \geq 2$ such that $\mathbb{H
\cap G}(n;\theta) (S)$ is connected while $S$ is isolated in
$\mathbb{H \cap G} (n;\theta)$. This is captured by the inclusion
\begin{equation}
C_n(\theta)^c \cap I(n;\theta)\ \subseteq  \bigcup_{S \subseteq
\mathcal{N}: ~ |S| \geq 2} ~ A_n (\theta ; S) \label{eq:BasicIdea}
\end{equation}
A moment of reflection should convince the reader that this union
need only be taken over all subsets $S$ of $\{1, \ldots , n \}$
with $2 \leq |S| \leq \lfloor \frac{n}{2} \rfloor $. A standard
union bound argument immediately gives
\begin{eqnarray}
\bP{ C_n(\theta)^c \cap I(n;\theta) } &\leq & \sum_{ S \subseteq
\mathcal{N}: 2 \leq |S| \leq \lfloor \frac{n}{2} \rfloor } \bP{
A_n (\theta ; S) }
\nonumber \\
&=& \sum_{r=2}^{ \lfloor \frac{n} {2} \rfloor } \left ( \sum_{S
\in \mathcal{N}_{n,r} } \bP{ A_n (\theta ; S) } \right )
\label{eq:BasicIdea+UnionBound}
\end{eqnarray}
where $\mathcal{N}_{n,r} $ denotes the collection of all subsets
of $\{ 1, \ldots , n \}$ with exactly $r$ elements.

For each $r=1, \ldots , n$, we simplify the notation by writing
$A_{n,r} (\theta) := A_n (\theta ; \{ 1, \ldots , r \} )$,
$B_{n,r} (\theta) := B_n (\theta ; \{ 1, \ldots , r \} )$ and
$C_{n,r}(\theta) := C_n (\theta ; \{ 1, \ldots , r \} )$. With a
slight abuse of notation, we use $C_n(\theta)$ for $r=n$ as
defined before. Under the enforced assumptions, exchangeability
yields
\[
\bP{ A_n (\theta ; S) } = \bP{ A_{n,r} (\theta ) }, \quad S \in
\mathcal{N}_{n,r}
\]
and the expression
\begin{equation}
\sum_{S \in \mathcal{N}_{n,r} } \bP{ A_n (\theta ; S) } = {n
\choose r} ~ \bP{ A_{n,r}(\theta ) } \label{eq:ForEach=r}
\end{equation}
follows since $|\mathcal{N}_{n,r} | = {n \choose r}$. Substituting
into (\ref{eq:BasicIdea+UnionBound}) we obtain the key bound
\begin{equation}
\bP{ C_n(\theta)^c \cap I(n;\theta) } \leq \sum_{r=2}^{ \lfloor
\frac{n}{2} \rfloor } {n \choose r} ~ \bP{ A_{n,r}(\theta ) } .
\label{eq:BasicIdea+UnionBound2}
\end{equation}

Consider a scaling $\theta: \mathbb{N}_0 \rightarrow \mathbb{N}_0
\times (0,1)$ as in the statement of Proposition
\ref{prop:OneLawAfterReduction}. Substitute $\theta$ by $\theta_n$
by means of this scaling in the right hand side of
(\ref{eq:BasicIdea+UnionBound2}). The proof of Proposition
\ref{prop:OneLawAfterReduction} will be completed once we show
\begin{equation}
\lim_{n \rightarrow \infty} \sum_{r=2}^{ \lfloor \frac{n}{2}
\rfloor } {n \choose r} ~ \bP{ A_{n,r}(\theta_n) } = 0.
\label{eq:OneLawToShow}
\end{equation}
The means to do so are provided in the next section.

\section{Bounding probabilities}
\label{sec:BoundingProbabilities}

Fix $n=2,3, \ldots $ and consider $\theta = (K,p)$ with $p$ in
$(0,1)$ and positive integer $K$ such that $K < n$.

\subsection{Bounding the probabilities $\bP{B_{n,r}(\theta)}$}
\label{subsec:BoundingProbabilitiesB}

The following result will be used to efficiently bound the
probability $\bP{B_{n,r}(\theta)}$.

\begin{lemma}
{\sl For each $r=2, \ldots , n-1$, we have the inequality
\begin{eqnarray}
\lefteqn{ \bP{ B_{n,r}(\theta) \Big | \Gamma_{n,1}, \ldots ,
\Gamma_{n,r} } } & &
\label{eq:BoundOnConditionalProbabilityB} \\
&\leq& \left ( 1-p \right )^{E^\star_{n,r}} \cdot
u_n(\theta)^{r(n-r) - E^\star_{n,r}} \nonumber
\end{eqnarray}
with $u_n(\theta)$ defined by (\ref{eq:Compute1}) and the rv
$E^\star_{n,r}$ given by
\begin{equation}
E^\star_{n,r} := \sum_{i=r+1}^{n} \sum_{\ell=1}^{r} \1{\ell \in
\Gamma_{n,i} }. \label{eq:E^Star}
\end{equation}
} \label{lem:BoundOnConditionalProbabilityB}
\end{lemma}

A proof of Lemma \ref{lem:BoundOnConditionalProbabilityB} is
available in Appendix \ref{App:B}. The rv $E^\star_{n,r}$, which
appears prominently in (\ref{eq:BoundOnConditionalProbabilityB}),
has a tail controlled through the following result.

\begin{lemma}
{\sl Fix $r=2, \ldots , n-1$. For any $t$ in $(0,1)$ we have
\begin{equation}
\bP{ E^\star_{n,r} \leq (1-t)  r K \cdot \frac{n-r}{n-1}  } \leq
e^{ - \frac{t^2}{2} r K \cdot \frac{n-r}{n-1} } .
\label{eq:bound_E_r_prime}
\end{equation}
} \label{lem:bound_E_r_prime}
\end{lemma}

\myproof Fix $n=2,3, \ldots $ and consider a positive integer $K$
such that $K < n$. From the facts reported in Section
\ref{sec:NegativeDependence}, the negative association of the rvs
(\ref{eq:NegativeAssociation2}) implies that of the rvs $\{ \1{
\ell \in \Gamma_{n,i} }, \ i=r+1, \ldots , n ; \ \ell=1, \ldots ,
r \}$. We are now in position to apply the Chernoff-Hoeffding
bound to the sum (\ref{eq:E^Star}). We use the bound in the form
\begin{equation}
\bP{ E^\star_{n,r} \leq (1-t) \bE{ E^\star_{n,r} }  } \leq
e^{-\frac{t^2}{2} \bE{ E^\star_{n,r} }  }
\end{equation}
as given for negatively associated rvs in \cite[Thm. 1.1, p.
6]{DubhashiPanconesi}. The conclusion (\ref{eq:bound_E_r_prime})
follows upon noting that
\[
\bE{ E^\star_{n,r} } = \sum_{i=r+1}^{n} \sum_{\ell=1}^{r} \bP{
\ell \in \Gamma_{n,i} } = r(n-r) \frac{K}{n-1}
\]
as we use (\ref{eq:ProbabilityBasic1}). \myendpf

\subsection{Bounding the probabilities $\bP{C_{n,r}(\theta)}$}
\label{subsec:BoundingProbabilitiesC}

For each $r=2, \ldots ,n$, let $\mathbb{H \cap G}_r(n;\theta)$
stand for the subgraph $\mathbb{H \cap G}(n;\theta)(S)$ when $S =
\{1 , \ldots , r \}$. Also let ${\cal T}_{r}$ denote the
collection of all spanning trees on the vertex set $\{1, \ldots ,
r \}$.

\begin{lemma}
{\sl Fix $r=2, \ldots , n$. For each $T$ in ${\cal T}_{r}$, we
have
\begin{equation}
\bP{ T \subset \mathbb{H \cap G}_r(n;\theta) } \leq \left ( p
\lambda_n(K) \right )^{r-1} \label{eq:ProbabilityOfTree}
\end{equation}
where the notation $T \subset \mathbb{H \cap G}_r(n;\theta)$
indicates that the tree $T$ is a subgraph spanning $\mathbb{H \cap
G}_r(n;\theta)$. } \label{lem:ProbabilityOfTree}
\end{lemma}

Since $ p \lambda_n(K)$ is the probability of link assignment, the
situation is reminiscent to the one found in ER graphs
\cite{Bollobas} and random key graphs \cite{YaganMakowskiISIT2009}
where in each case the bound (\ref{eq:ProbabilityOfTree}) holds
with equality.

\myproof Fix $r=2, 3, \ldots, n$ and pick a tree $T$ in ${\cal
T}_{r}$. Let ${\cal E}(T)$ be the set of edges that appear in $T$.
It is plain that $T \subseteq \mathbb{H \cap G}_{r} (n,; \theta )
$ occurs if and only if the set of conditions
\[
\begin{array}{c}
\Sigma_{n,i} \cap \Sigma_{n,j} \neq \emptyset  \\
\mbox{and} \\
B_{ij} (p) =1 \\
\end{array}
, \quad \{i,j\} \in {\cal E}(T)
\]
holds. Therefore, under the enforced independence assumptions,
since $| {\cal E}(T) | = r-1$, we get
\begin{eqnarray}
\lefteqn{ \bP{ T \subset \mathbb{H \cap G}_r(n;\theta) } } & &
\nonumber \\
&=& p^{r-1} \cdot \bE{ \prod_{i,j : \{i,j\} \in {\cal E}(T) } \1 {
\Sigma_{n,i} \cap \Sigma_{n,j} \neq \emptyset } }
\nonumber \\
&=& p ^ {r-1} \cdot \bE{ \prod_{i,j : \{i,j\} \in {\cal E}(T) }
\1{ i \in \Gamma_{n,j} \: \vee \: j \in \Gamma_{n,i} } }
\nonumber \\
&\leq& p ^ {r-1} \cdot \prod_{i,j : \{i,j\} \in {\cal E}(T) } \bP{
i \in \Gamma_{n,j} \: \vee \: j \in \Gamma_{n,i} }
\end{eqnarray}
by making use of (\ref{eq:KeyConsequenceForProbabilities}) with
the negatively associated rvs (\ref{eq:NegativeAssociation3}). The
desired result (\ref{eq:ProbabilityOfTree}) is now immediate from
(\ref{eq:LinkAssignmentinH}) and the relation $|{\cal E}(T)| =
r-1$. \myendpf

As in ER graphs \cite{Bollobas} and random key graphs
\cite{YaganMakowskiISIT2009} we have to the following bound.

\begin{lemma}
{\sl For each $r=2, \ldots , n$, we have
\begin{equation}
\bP{ C_{n,r}(\theta) } \leq r^{r-2} \left ( p \lambda_n(K)
\right)^{r-1} . \label{eq:ProbabilityOfC}
\end{equation}
} \label{lem:ProbabilityOfC}
\end{lemma}

\myproof Fix $r=2, \ldots , n$. If $\mathbb{H \cap G}_r
(n;\theta)$ is a connected graph, then it must contain a spanning
tree on the vertex set $\{1, \ldots . r \}$, and a union bound
argument yields
\begin{equation}
\bP{ C_{n,r}(\theta) } \leq \sum_{T \in {\cal T}_r } \bP{ T
\subset \mathbb{H \cap G}(n;\theta)(S) } .
\label{eq:ProbabilityOfAwithTrees}
\end{equation}
By Cayley's formula \cite{Martin} there are $r^{r-2}$ trees on $r$
vertices, i.e., $| {\cal T}_{r}| = r^{r-2}$, and
(\ref{eq:ProbabilityOfC}) follows upon making use of
(\ref{eq:ProbabilityOfTree}). \myendpf

\section{A proof of Proposition \ref{prop:OneLawAfterReduction}
         (Part II)}
\label{sec:ProofConnectivityII}

Consider a scaling $\theta: \mathbb{N}_0 \rightarrow \mathbb{N}_0
\times (0,1)$ as in the statement of Proposition
\ref{prop:OneLawAfterReduction}. Pick integers $R \geq 2$ and
$n^\star (R) \geq 2(R+1)$ (to be specified in Section
\ref{sec:ToShowSecondPiece2}). On the range $n \geq n^\star(R)$ we
consider the decomposition
\begin{eqnarray}
\lefteqn{ \sum_{r=2}^{\lfloor \frac{n}{2} \rfloor} {n \choose r} ~
\bP{ A_{n,r} (\theta_n) }} &&
\nonumber \\
&=& \sum_{r=2}^{ R } {n \choose r} ~ \bP{ A_{n,r} (\theta_n) } +
\sum_{ r=R+1}^{\lfloor \frac{n}{2} \rfloor} { n \choose r} ~ \bP{
A_{n,r} (\theta_n) } , \nonumber \label{eq:AnotherDecomposition}
\end{eqnarray}
and let $n$ go to infinity. The desired convergence
(\ref{eq:OneLawToShow}) will be established if we show
\begin{equation}
\lim_{n \rightarrow \infty} { n \choose r} ~ \bP{ A_{n,r}
(\theta_n) } = 0 \label{eq:StillToShow0}
\end{equation}
for {\em each} $r=2,3, \ldots $ and
\begin{equation}
\lim_{n \rightarrow \infty} \sum_{ r=R+1}^{\lfloor \frac{n}{2}
\rfloor} { n \choose r} ~ ~ \bP{ A_{n,r} (\theta_n) } = 0.
\label{eq:StillToShow2}
\end{equation}

We establish (\ref{eq:StillToShow0}) and (\ref{eq:StillToShow2})
in turn. Throughout, we make use of the standard bounds
\begin{equation}
{n \choose r} \leq \left ( \frac{e n}{r} \right )^r, \quad r=1,
\ldots , n \label{eq:CombinatorialBound1}
\end{equation}
for each $n=2,3, \ldots$.

\subsection{Establishing (\ref{eq:StillToShow0})}
\label{sec:ToShowSecondPiece0}

Fix $r =2,3, \ldots $ and consider $n=2,3, \ldots $ such that $ r
< n$. Also let $\theta = (K,p)$ with $p$ in $(0,1)$ and positive
integer $K$ such that $K < n$. With (\ref{eq:E^Star}) in mind, for
each $i=1, \ldots , r$, we note that
\begin{eqnarray}
\sum_{k=r+1}^n \1{ k \in \Gamma_{n,i} } &=& \sum_{k=1}^n \1{ k \in
\Gamma_{n,i} } - \sum_{k=1}^r \1{ k \in \Gamma_{n,i} }
\nonumber \\
&=& K - \sum_{k=1}^r \1{ k \in \Gamma_{n,i} }
\end{eqnarray}
since $|\Gamma_{n,i}| = K$. The bounds
\[
(K-r)^+ \leq \sum_{k=r+1}^n \1{ k \in \Gamma_{n,i} } \leq K
\]
follow, whence
\[
r(K-r)^+ \leq E^\star_{n,r} \leq r K.
\]
It is also the case that
\[
r(n-r-K)^+ \leq r(n-r) - E^\star_{n,r}.
\]

Reporting these lower bounds into
(\ref{eq:BoundOnConditionalProbabilityB}), we get
\begin{eqnarray}
\lefteqn{ \bP{ B_{n,r}(\theta) \Big | \Gamma_{n,1}, \ldots ,
\Gamma_{n,r} } } & &
\nonumber \\
&\leq& \left ( 1-p \right )^{ r(K-r)^+ } \cdot u_n(\theta)^{ r(
n-r-K )^+ }
\label{eq:A} \\
&\leq& \left ( 1-p \right )^{ r(K-r) } \cdot u_n(\theta)^{ r(
n-r-K ) } \nonumber
\end{eqnarray}
since $0 < p, u_n(\theta) < 1$. If we set
\[
F_{n,r} (\theta) := \left ( 1-p \right )^{ (K-r) } \cdot
u_n(\theta)^{ (n-r-K) },
\]
it is now plain that
\begin{eqnarray}
\lefteqn{ \bP{ A_{n,r}(\theta) } } & &
\nonumber \\
&=& \bE{ \1{ C_{n,r}(\theta) } \bP{ B_{n,r}(\theta) \Big |
\Gamma_{n,1}, \ldots , \Gamma_{n,r} } }
\nonumber \\
&\leq& \bP{ C_{n,r}(\theta) } \cdot F_{n,r} (\theta)^r .
\label{eq:B}
\end{eqnarray}
Applying Lemma \ref{lem:ProbabilityOfC} we find
\begin{eqnarray}
\lefteqn{ {n \choose r }\bP{ A_{n,r}(\theta) } } & &
\nonumber \\
&\leq& {n \choose r } \bP{ C_{n,r}(\theta) } \cdot F_{n,r}
(\theta)^r
\nonumber \\
&\leq& \left ( \frac{ en }{r} \right )^r r^{r-2} \left ( p
\lambda_n(K) \right )^{r-1} F_{n,r} (\theta)^r
\nonumber \\
&=& \frac{1}{r^2} \left ( en \right )^r \left ( p \lambda_n(K)
\right )^{r-1} F_{n,r} (\theta)^r \label{eq:X}
\end{eqnarray}
as we make use of (\ref{eq:CombinatorialBound1}).

We also note that
\begin{equation}
F_{n,r}(\theta) \leq e^{ F^\star_{n,r}(\theta)} \label{eq:ZZ}
\end{equation}
with
\begin{eqnarray}
\lefteqn{ F^\star_{n,r}(\theta)} & &
\nonumber \\
&:=& (K-r) \log (1-p)  - ( n-r-K ) p \frac{K}{n-1}
\nonumber \\
&=& (K-r) \log (1-p) - \left ( 1 - \frac{ K}{n-1}  - \frac{
r-1}{n-1} \right ) p K
\nonumber \\
&=& (K-r) \log (1-p) - p \left ( K - \frac{K^2}{n-1} \right ) +
\frac{ r-1}{n-1} p K
\nonumber \\
&=& K \left ( p + \log (1-p) \right ) - r \log (1-p)
\nonumber \\
& & \quad -~ p \left (2 K - \frac{K^2}{n-1} \right ) + \frac{
r-1}{n-1} p K . \label{eq:ZZZ}
\end{eqnarray}

Now, pick any given positive integer $r=2,3, \ldots$ and consider
a scaling $\theta: \mathbb{N}_0 \rightarrow \mathbb{N}_0 \times
(0,1)$ such that $\lim_{n \to \infty}p_n=p^\star$ exists and
(\ref{eq:scalinglaw}) holds for some $c>\tau (p ^ \star)$. Replace
$\theta$ by $\theta_n$ in (\ref{eq:X}) according to this scaling.
In order to establish (\ref{eq:StillToShow0}) it suffices to show
that
\begin{equation}
\lim_{n \rightarrow \infty} \left ( en \right )^r \left ( p_n
\lambda_n(K_n) \right )^{r-1} \cdot F_{n,r} (\theta_n)^r = 0 .
\label{eq:XX}
\end{equation}

For $n$ sufficiently large, from (\ref{eq:scalinglawEquivalent})
and (\ref{eq:X}) we first get
\begin{eqnarray}
\lefteqn{ {n \choose r }\bP{ A_{n,r}(\theta) } } & &
\nonumber \\
&\leq& \left ( en \right )^r \left ( p_n \lambda_n(K_n)
\right)^{r-1} \cdot F_{n,r} (\theta_n)^r
\nonumber \\
&=& \left ( en \right )^r \left ( c_n \frac{\log n }{n-1} \right
)^{r-1} \cdot F_{n,r} (\theta_n)^r
\nonumber \\
&=& en \left ( ec_n \frac{n}{n-1} \log n \right )^{r-1} \cdot
F_{n,r} (\theta_n)^r . \label{eq:QQ}
\end{eqnarray}

On the other hand, upon making use of the bounds at
(\ref{eq:usefulinequality}), we find
\begin{eqnarray}
F^\star_{n,r}(\theta_n ) &\leq & K_n \left ( p_n + \log (1-p_n)
\right ) - r \log (1-p_n)
\nonumber \\
& & -~ p_n \left (2 K_n - \frac{K_n^2}{n-1} \right ) + \frac{r}{n}
~ p_n K_n
\nonumber \\
&=& K_n \left ( p_n + \log (1-p_n) \right ) - r \log (1-p_n)
\nonumber \\
& & -~ c_n \log n + \frac{r}{n} ~ p_n K_n
\nonumber \\
&\leq& K_n \left ( p_n + \log (1-p_n) \right )  - c_n \log n
\nonumber \\
& & - r \log (1-p_n) + \frac{r}{n} ~ c_n \log n
\nonumber \\
&=& p_n K_n \left ( 1 + \frac{\log (1-p_n)}{p_n} \right ) - c_n
\log n
\nonumber \\
& & - ~r \log (1-p_n) + \frac{r}{n} ~ c_n \log n
\nonumber \\
&\leq& \frac{c_n}{2} \log n \cdot \left ( 1 + \frac{\log
(1-p_n)}{p_n} \right ) - c_n \log n
\nonumber \\
& & -~ r \log (1-p_n) + \frac{r}{n} ~ c_n \log n
\nonumber \\
&=& - \frac{c_n}{2} \cdot \left ( 1 - \frac{\log (1-p_n)}{p_n}
\right ) \log n
\nonumber \\
& & -~ r \log (1-p_n) + \frac{r}{n} ~ c_n \log n .
\nonumber \\
&=& \log n \left( -\frac{c_n-\frac{2rp_n}{\log n}}{2} \left( 1 -
\frac{\log(1-p_n)}{p_n}\right) \right)
\nonumber \\
& & -~ r p_n + \frac{r}{n} ~ c_n \log n
\nonumber \\
&\leq& - \frac{\log n}{2} \left( c_n-\frac{2rp_n}{\log n} \right )
\left( 1 - \frac{\log(1-p_n)}{p_n}\right)
\nonumber \\
& & +~ \frac{r}{n} ~ c_n \log n . \label{eq:RR}
\end{eqnarray}

As a result, (\ref{eq:ZZZ}) implies
\begin{eqnarray}
\lefteqn{ n F_{n,r}(\theta_n)^r } & &
\label{eq:TT} \\
&\leq& n^{1- \frac{r}{2} \left ( c_n-\frac{2rp_n}{\log n} \right )
\cdot \left ( 1 - \frac{\log (1-p_n)}{p_n} \right ) }  e^{ o(1) }
. \nonumber
\end{eqnarray}
Under the enforced assumptions of Theorem
\ref{thm:OneLaw+Connectivity} we get
\begin{eqnarray}
\lefteqn{\lim_{n \rightarrow \infty} \left ( 1 - \frac{r}{2} \left
( c_n-\frac{2rp_n}{\log n} \right ) \cdot \left ( 1 - \frac{\log
(1-p_n)}{p_n} \right ) \right ) } & &
\nonumber \\
&=& 1- r \frac{c}{2} \cdot \left ( 1 - \frac{\log
(1-p^\star)}{p^\star} \right )\hspace{1cm}
\nonumber \\
&=& 1 - r \frac{c}{\tau(p^\star)} < 0,
\end{eqnarray}
and the desired conclusion (\ref{eq:XX}) follows upon making use
of the inequalities (\ref{eq:QQ}) and (\ref{eq:TT}).

\subsection{Establishing (\ref{eq:StillToShow2})}
\label{sec:ToShowSecondPiece2}

Fix $n=2,3, \ldots $ and consider $\theta = (K,p)$ with $p$ in
$(0,1)$, and positive integer $K$ such that $K < n$.

Pick $r=1,2, \ldots , n-1$. By Lemma
\ref{lem:BoundOnConditionalProbabilityB} we conclude that
\begin{equation}
\bP{ B_{n,r}(\theta) \Big | \Gamma_{n,1}, \ldots , \Gamma_{n,r} }
\leq (1-p)^{E^\star_{n,r} }
\end{equation}
since $0 < u_n(\theta) < 1$, and preconditioning arguments similar
to the ones leading to (\ref{eq:B}) yield
\[
\bP{ A_{n,r}(\theta) } \leq \bE{ \1{ C_{n,r}(\theta) }
(1-p)^{E^\star_{n,r} } }.
\]
The event $C_{n,r}(\theta)$ depends only on $\Gamma_{n,1}, \ldots,
\Gamma_{n,r}$ whereas $E^\star_{n,r}$ is determined solely by
$\Gamma_{n,r+1}, \ldots, \Gamma_{n,n}$. Thus, the event
$C_{n,r}(\theta)$ is independent of the rv $\left( 1- p \right
)^{E^\star_{n,r}}$ under the enforced assumptions, whence
\begin{equation}
\bP{ A_{n,r}(\theta) } \leq \bP{ C_{n,r}(\theta) } \bE{
(1-p)^{E^\star_{n,r} } }. \label{eq:CC}
\end{equation}

Pick $t$ arbitrary in $(0,1)$ and recall
 Lemma \ref{lem:bound_E_r_prime}.
A simple decomposition argument shows that
\begin{eqnarray}
\lefteqn{ \bE{\left( 1- p \right )^{E^\star_{n,r}} } } & &
\nonumber \\
&\leq& \bE{\left( 1- p \right )^{E^\star_{n,r} } \1{ E^\star_{n,r}
> ( 1 - t ) r K \cdot \frac{ n - r }{ n - 1 } }  }
\nonumber \\
& & ~ + \bP{ E^\star_{n,r } \leq ( 1 - t ) r K \cdot \frac{ n - r
} { n - 1 } }
\nonumber \\
&\leq& \left( 1- p \right )^{( 1 - t ) r K \cdot \frac{n - r}{n -
1} } + e^{-\frac{t^2}{2} r K \cdot \frac{n-r}{n-1} }
\nonumber \\
&\leq& e^{- ( 1 - t ) rp K \cdot \frac{n - r}{n - 1} } +
e^{-\frac{t^2}{2} r K \cdot \frac{n-r}{n-1} }
\nonumber \\
&\leq& e^{- ( 1 - t ) rp K \cdot \frac{n - r}{n - 1} } +
e^{-\frac{t^2}{2} rp K \cdot \frac{n-r}{n-1} } . \nonumber
\end{eqnarray}
Therefore, whenever $r=2, 3, \ldots, \lfloor \frac{n}{2} \rfloor$,
we have
\begin{equation}
\bE{\left( 1- p \right )^{E^\star_{n,r}} } \leq e^{- \frac{1-t}{2}
\cdot  r p K } + e^{-\frac{t^2}{4} \cdot r p K }
\label{eq:noN_fixed_r_int_step}
\end{equation}
since on that range we have
\[
\frac{ n - r }{ n - 1 } \geq \frac{n/2}{n-1}\geq \frac{1}{2}.
\]

Now consider a scaling $\theta: \mathbb{N}_0 \rightarrow
\mathbb{N}_0 \times (0,1)$ such that $\lim_{n \to
\infty}p_n=p^\star$ exists and (\ref{eq:scalinglaw}) holds for
some $c>\tau (p ^ \star)$. Replace $\theta$ by $\theta_n$ in both
(\ref{eq:CC}) and (\ref{eq:noN_fixed_r_int_step}) according to
this scaling and use the bound of Lemma \ref{lem:ProbabilityOfC}
in the resulting inequalities. Pick an integer $R \geq 2$ (to be
further specified shortly) and for $n \geq 2(R+1)$ note that
\begin{eqnarray}
\lefteqn{\sum_{r=R+1}^{\lfloor \frac{n}{2} \rfloor} { n \choose r}
~ ~ \bP{ A_{n,r} (\theta_n) } } & &
\nonumber \\
&\leq& \sum_{r=R+1}^{\lfloor \frac{n}{2} \rfloor} { n \choose r}
r^{r-2} \left ( p_n \lambda_n(K_n) \right )^{r-1} e^{ - \frac{1 -
t}{2} \cdot r p_n K_n }
\nonumber \\
& & ~ + \sum_{r=R+1}^{\lfloor \frac{n}{2} \rfloor} { n \choose r}
r^{r-2} \left ( p_n \lambda_n(K_n) \right )^{r-1}
e^{-\frac{t^2}{4} \cdot r p_n K_n }
\nonumber \\
&\leq& \sum_{r=R+1}^{\lfloor \frac{n}{2} \rfloor} e n \left ( ec_n
\frac{n}{n-1} \log n \right )^{r-1} e^{ - \frac{1 - t}{2} \cdot r
p_n K_n }
\nonumber \\
& & ~+ \sum_{r=R+1}^{\lfloor \frac{n}{2} \rfloor} e n \left ( e
c_n \frac{n}{n-1} \log n \right )^{r-1} e^{-\frac{t^2}{4} \cdot r
p_n K_n }
\nonumber
\end{eqnarray}
by the same arguments as the ones leading to (\ref{eq:QQ}). Upon
invoking the lower bound in (\ref{eq:usefulinequality}) we now
conclude for all sufficiently large $n > 2(R+1)$ that
\begin{eqnarray}
\lefteqn{\sum_{r=R+1}^{\lfloor \frac{n}{2} \rfloor} { n \choose r}
~ ~ \bP{ A_{n,r} (\theta_n) } } & &
\nonumber \\
&\leq& \sum_{r=R+1}^{\lfloor \frac{n}{2} \rfloor} e n \left ( ec_n
\frac{n}{n-1} \log n \right )^{r} e^{ - \frac{1 - t}{4} \cdot r
c_n \log n }
\nonumber \\
& & ~+ \sum_{r=R+1}^{\lfloor \frac{n}{2} \rfloor} e n \left ( ec_n
\frac{n}{n-1} \log n \right )^{r} e^{-\frac{t^2}{8} \cdot r c_n
\log n } .
\nonumber \\
&\leq& \sum_{r=R+1}^{\infty} e n \left ( ec_n \frac{n}{n-1} \log n
\cdot n^{ - \frac{1 - t}{4} \cdot c_n } \right )^{r}
\nonumber \\
& & ~+ \sum_{r=R+1}^{\infty} e n \left ( ec_n \frac{n}{n-1} \log n
\cdot n^{-\frac{t^2}{8} \cdot c_n } \right )^{r} .
\nonumber
\end{eqnarray}
Furthermore, for all sufficiently large $n \geq 2 (R+1)$ it also
the case that
\begin{equation}
e c_n \frac{n}{n-1} \log n \cdot \max \left ( n^{ - \frac{1-t}{4}
c_n }, n^{ - \frac{t^2}{8} c_n } \right ) < 1 \label{eq:MaxHolds}
\end{equation}
and the two infinite series converge. Let $n^\star(R)$ denote any
integer larger than $2(R+1)$ such that (\ref{eq:MaxHolds}) holds
for all $n \geq n^\star(R)$. On that range, by our earlier
discussion we get
\[
\sum_{ r=R+1}^{\lfloor \frac{n}{2} \rfloor} { n \choose r} ~ \bP{
A_{n,r} (\theta_n) } \leq e \left ( e c_n \frac{n}{n-1} \log n
\right )^{R + 1} \left ( \ldots \right )
\]
with
\begin{eqnarray}
\ldots &:=&
 \frac{ n^{1 - \frac{1-t}{4} c_n(R+1)} }
      { 1 - ec_n \frac{n}{n-1} \log n \cdot n^{- \frac{1 - t}{4} c_n } }
\nonumber \\
& & ~ + \frac{ n^{1 - \frac{t^2}{8} c_n(R+1)} }
         {1 - ec_n \frac{n}{n-1} \log n \cdot n^{- \frac{t^2}{8} c_n } } .
\nonumber
\end{eqnarray}

Finally, let $n$ go to infinity in this last expression: The
desired conclusion (\ref{eq:StillToShow2}) follows whenever the
conditions $(1-t)c(R+1) > 4$ and $c(R+1)t^2 > 8$ are satisfied.
This can be achieved by taking $R$ so that
\[
R+1 > \max \left ( \frac{4}{c (1-t)} , \frac{8}{c t^ 2} \right ).
\]
This is always feasible for any given $t $ in $(0,1)$ by taking
$R$ sufficiently large. \myendpf

\appendices

\section{A proof of Proposition \ref{prop:EvalCrossMoment}}
\label{App:A}

The basis for deriving (\ref{eq:EvalCrossMoment}) lies in the
observation that nodes $1$ and $2$ are both isolated in $\mathbb{H
\cap G}(n;\theta)$ if and only if each edge in $\mathbb{H}(n;K)$
incident to one of these nodes is {\em not} present in
$\mathbb{G}(n;p)$. Thus, $\chi_{n,1} (\theta) = \chi_{n,2}
(\theta) =1$ if and only if both sets of conditions
\[
B_{1j}(p) = 0 \quad \mbox{if} \quad \Sigma_{n,1} \cap \Sigma_{n,j}
\neq \emptyset , \ j \in {\cal N}_{-1}
\]
and
\[
B_{2k} (p) = 0 \quad \mbox{if} \quad \Sigma_{n,2} \cap
\Sigma_{n,k} \neq \emptyset , \ k \in {\cal N}_{-2}
\]
hold.

To formalize this observation, we introduce the random sets
$N_{n,1}(\theta)$ and $N_{n,2}(\theta)$ defined by
\begin{equation}
N_{n,1}(\theta) := \{ j =3, \ldots , n : \ j \in \Gamma_{n,1} \
\vee 1 \in \Gamma_{n,j} \} \label{eq:definitioN_N_1}
\end{equation}
and
\begin{equation}
N_{n,2}(\theta) := \{ k =3, \ldots , n : \ k \in \Gamma_{n,2} \
\vee 2 \in \Gamma_{n,k} \} . \label{eq:definitioN_N_1}
\end{equation}
Thus, node $j$ in $N_{n,1}(\theta)$ is neither node $1$ nor node
$2$, and is K-adjacent to node $1$. Similarly, node $k$ in
$N_{n,2}(\theta)$ is neither node $1$ nor node $2$, and is
K-adjacent to node $2$. Let $Z_n(\theta)$ denote the total number
of edges in $\mathbb{H}(n;K)$ which are incident to either node
$1$ or node $2$. It is plain that
\begin{eqnarray}
Z_n (\theta) &=& |N_{n,1}(\theta)| + |N_{n,2}(\theta)|
\nonumber \\
& &  ~ + \1{ 2 \in \Gamma_{n,1} \ \vee 1 \in \Gamma_{n,2} }
\label{eq:Z}
\end{eqnarray}
with the last term accounting for the possibility that nodes $1$
and $2$ are K-adjacent. By conditioning on the rvs $\Gamma_{n,1},
\ldots , \Gamma_{n,n}$, we readily conclude that
\begin{equation}
\bE{ \chi_{n,1} (\theta) \chi_{n,2} (\theta) } = \bE{
(1-p)^{Z_n(\theta)} } \label{eq:BasicEquality}
\end{equation}
under the enforced independence of the collections of rvs $\{
\Gamma_{n,1}, \ldots , \Gamma_{n,n} \}$ and $\{B_{ij}(p), 1 \leq i
< j \leq n\}$.

To proceed we need to assess the various contributions to
$Z_n(\theta)$: Using (\ref{eq:BasicSetIdentity}) we find
\begin{eqnarray}
|N_{n,1}(\theta)| &=& \sum_{j=3}^n \1{ j \in \Gamma_{n,1} \ \vee 1
\in \Gamma_{n,j} }
\nonumber \\
&=& \sum_{j=3}^n \1{ j \in \Gamma_{n,1} }
  + \sum_{j=3}^n \1{ 1 \in \Gamma_{n,j} }
\nonumber \\
& &
  - \sum_{j=3}^n \1{ j \in \Gamma_{n,1}, \ 1 \in \Gamma_{n,j} }
\nonumber \\
&=& \sum_{j=3}^n \1{ j \in \Gamma_{n,1} } + \sum_{j=3}^n \1{ j
\not \in \Gamma_{n,1}, \ 1 \in \Gamma_{n,j} }
\nonumber \\
&=& K - \1{ 2 \in \Gamma_{n,1} }
\nonumber \\
& & + \sum_{j=3}^n \1{ j \not \in \Gamma_{n,1}, \ 1 \in
\Gamma_{n,j} }
\end{eqnarray}
where the last step used the fact $|\Gamma_{n,1}|=K$. Similar
arguments show that
\begin{eqnarray}
|N_{n,2}(\theta)| &=& \sum_{k=3}^n \1{ k \in \Gamma_{n,2} \ \vee 2
\in \Gamma_{n,k} }
\nonumber \\
&=& K - \1{ 1 \in \Gamma_{n,2} }
\nonumber \\
& & + \sum_{k=3}^n \1{ k \not \in \Gamma_{n,2}, \ 2 \in
\Gamma_{n,k} }.
\end{eqnarray}

As a result, from the definition of $Z_n(\theta)$ we get
\begin{equation}
Z_n(\theta) = 2K - \1{ 2 \in \Gamma_{n,1}, \ 1 \in \Gamma_{n,2} }
+ Z^\star_n(\theta) \label{eq:Z_n}
\end{equation}
upon using (\ref{eq:BasicSetIdentity}) one more time, where
\begin{eqnarray}
Z^\star_n(\theta) &:=& \sum_{j=3}^n \1{ j \not \in \Gamma_{n,1}, \
1 \in \Gamma_{n,j} }
\nonumber \\
& & + \sum_{j=3}^n \1{ j \not \in \Gamma_{n,2}, \ 2 \in
\Gamma_{n,j} }.
\end{eqnarray}

In order to evaluate the expression (\ref{eq:BasicEquality}), we
first compute the conditional expectation
\begin{equation}
\bE{ (1-p)^{Z_n(\theta)} \Big | \Gamma_{n,1}, \  \Gamma_{n,2} }.
\label{eq:ConditionalExpectation}
\end{equation}
From (\ref{eq:Z_n}) we see that this quantity can be evaluated as
the product of the two terms
\begin{equation}
(1-p)^{2K - \left (  \1{ 2 \in \Gamma_{n,1}, 1 \in \Gamma_{n,2} }
\right ) } \label{eq:Factor1}
\end{equation}
and
\begin{equation}
\bE{ (1-p)^{ Z^\star_n(\theta) } \Big | \Gamma_{n,1}, \
\Gamma_{n,2} }. \label{eq:Factor2}
\end{equation}

To evaluate this last conditional expectation, for each $j=3,
\ldots , n$, we set
\begin{eqnarray}
\lefteqn{ V_{n,j} (\theta ; S, T) } & &
\nonumber \\
&:=& \bE{ (1-p)^ { \1{ j \not \in S, \ 1 \in \Gamma_{n,j} } + \1{
j \not \in T, \ 2 \in \Gamma_{n,j} } } } \nonumber
\end{eqnarray}
with $S$ and $T$ subsets of ${\cal N}$, each being of size $K$. It
is straightforward to check that
\begin{eqnarray}
\lefteqn{ V_{n,j} (\theta ; S, T) } & &
\nonumber \\
&=& \1{ j \not \in S } \1{ j \not \in T } \bE{ (1-p)^ { \1{ 1 \in
\Gamma_{n,j} } + \1{ 2 \in \Gamma_{n,j} } } }
\nonumber \\
& & + \1{ j \not \in S } \1{ j \in T } \bE{ (1-p)^{ \1{ 1 \in
\Gamma_{n,j} } } }
\nonumber \\
& & + \1{ j \not \in T } \1{ j \in S } \bE{ (1-p)^{ \1{ 2 \in
\Gamma_{n,j} } } }
\nonumber \\
& & + \1{ j \in S } \1{ j \in T }. \nonumber
\end{eqnarray}
Then, with the notation introduced earlier in Section
\ref{sec:NegativeDependence}, we can write
\begin{eqnarray}
\lefteqn{ V_{n,j} (\theta ; S, T) } & &
\nonumber \\
&=& \1{ j \not \in S } \1{ j \not \in T } b_n (\theta)
\nonumber \\
& & + \left ( \1{ j \not \in S } \1{ j \in T } + \1{ j \not \in T
} \1{ j \in S } \right ) u_n (\theta)
\nonumber \\
& & + \1{ j \in S } \1{ j \in T }. \nonumber
\end{eqnarray}

Next, the two rvs $\Gamma_{n,1}$ and $\Gamma_{n,2}$ being jointly
independent of the rvs $\Gamma_{n,3}, \ldots , \Gamma_{n,n}$, we
find
\begin{eqnarray}
\lefteqn{ \bE{ (1-p)^{ Z^\star_n(\theta) } \Big | \Gamma_{n,1}, \
\Gamma_{n,2} } } & &
\nonumber \\
&=& \prod_{j=3}^n V_{n,j} (\theta ; \Gamma_{n,1}, \Gamma_{n,2} )
\nonumber \\
&=& b_n(\theta)^{B_n(\theta)} \cdot u_n(\theta)^{U_n(\theta)}
\label{eq:ExpressionForConditionalExpectation}
\end{eqnarray}
where the rvs $B_n(\theta)$ and $U_n(\theta)$ are given by
(\ref{eq:B_n}) and (\ref{eq:U_n}), respectively. Therefore, since
\[
\bE{ (1-p)^{Z_n(\theta)} } = \bE{ \bE{ (1-p)^{Z_n(\theta) } \Big |
\Gamma_{n,1}, \Gamma_{n,2} } }
\]
by a standard preconditioning argument, we get the expression
(\ref{eq:EvalCrossMoment}) upon writing
(\ref{eq:ConditionalExpectation}) as the product of the quantities
(\ref{eq:Factor1}) and (\ref{eq:Factor2}), and using
(\ref{eq:ExpressionForConditionalExpectation}). \myendpf

\section{A proof of Lemma \ref{lem:BoundOnConditionalProbabilityB}}
\label{App:B}

The defining conditions for $B_{n,r}(\theta)$ lead to the
representation
\[
B_{n,r}(\theta) = \cap_{i=1}^r \cap_{k=r+1}^n E_{n,ik}(\theta)
\]
where we have set
\[
E_{n,ik}(\theta) := \left ( [ k \notin \Gamma_{n,i} ] \cap [ i
\notin \Gamma_{n,k} ] \right ) \cup [ B_{ik}(p) = 0 ]
\]
with $i=1, \ldots ,r$ and $k=r+1, \ldots , n$. In terms of
indicator functions, with the help of (\ref{eq:BasicSetIdentity})
this definition reads
\begin{eqnarray}
\lefteqn{ \1{ E_{n,ik}(\theta) } } & &
\nonumber \\
&=& \1{ k \notin \Gamma_{n,i} } \1{ i \notin \Gamma_{n,k} } +
(1-B_{ik}(p) )
\nonumber \\
& & - \1{ k \notin \Gamma_{n,i} } \1{ i \notin \Gamma_{n,k} }
(1-B_{ik}(p) )
\nonumber \\
&=& (1-B_{ik}(p) ) + \1{ k \notin \Gamma_{n,i} } \1{ i \notin
\Gamma_{n,k} } B_{ik}(p) . \nonumber
\end{eqnarray}
Therefore, under the enforced independence assumptions,
\begin{eqnarray}
\lefteqn{ \bP{ B_{n,r}(\theta) \Big | \Gamma_{n,1}, \ldots ,
\Gamma_{n,n} } } &&
\nonumber \\
&=& \bE{\prod_{i=1}^r \prod_{k=r+1}^n W \left ( \1{ k \notin
\Gamma_{n,i} } \1{ i \notin \Gamma_{n,k} }; p \right )} \nonumber
\end{eqnarray}
where
\[
W(x;p) = 1-p + px, \quad x \in \mathbb{R}.
\]
Since $W(x,p) = (1-p)^{1-x}$ for $x=0,1$, we obtain
\begin{eqnarray}
\lefteqn{ \bP{ B_{n,r}(\theta) \Big | \Gamma_{n,1}, \ldots ,
\Gamma_{n,n} } } & &
\nonumber \\
&=& \bE{\prod_{i=1}^r \prod_{k=r+1}^n \left ( 1 - p \right )^{1-
\1{ k \notin \Gamma_{n,i} } \1{ i \notin \Gamma_{n,k} } }},
\nonumber
\end{eqnarray}
and it is now plain that
\begin{eqnarray}
\lefteqn{ \bP{ B_{n,r}(\theta) \Big | \Gamma_{n,1}, \ldots ,
\Gamma_{n,r} } } & &
\nonumber \\
&=& (1-p)^{r(n-r)} G_{n,r} (\Gamma_{n,1}, \ldots , \Gamma_{n,r};
\theta )
\nonumber
\end{eqnarray}
where we have set
\begin{eqnarray}
\lefteqn{ G_{n,r} (S_1, \ldots , S_r; \theta) } & &
\nonumber \\
&=& \bE{ \prod_{i=1}^r \prod_{k=r+1}^n \left ( 1 - p \right )^ {-
\1{ k \notin S_{i} } \1{ i \notin \Gamma_{n,k} } } }
\nonumber
\end{eqnarray}
with $S_1, \ldots , S_r$ subsets of ${\cal N}$, each of size $K$.

Next, we find
\begin{eqnarray}
\lefteqn{ G_{n,r} (S_1, \ldots , S_r; \theta) } & &
\nonumber \\
&=& \bE{ \prod_{k=r+1}^n \prod_{i=1}^r \left ( 1 - p \right )^ {-
\1{ k \notin S_{i} } \1{ i \notin \Gamma_{n,k} } } }
\nonumber \\
&=& \bE{ \prod_{k=r+1}^n \left ( 1 - p \right )^ {- \sum_{i=1}^r
\1{ k \notin S_{i} } \1{ i \notin \Gamma_{n,k} } } }
\nonumber \\
&=& \prod_{k=r+1}^n \bE{ \left ( 1 - p \right )^ {- \sum_{i=1}^r
\1{ k \notin S_{i} } \1{ i \notin \Gamma_{n,k} } } }
\nonumber
\end{eqnarray}
as we again use the enforced independence assumptions. Fix $k=r+1,
\ldots , n$ and note that
\begin{eqnarray}
\lefteqn{ \bE{ \left ( 1 - p \right )^ {- \sum_{i=1}^r \1{ k
\notin S_{i} } \1{ i \notin \Gamma_{n,k} } } } } & &
\nonumber \\
&=& \bE{ \prod_{i=1}^r \left ( \left ( 1 - p \right )^{- \1{ k
\notin S_{i} }} \right )^{\1{ i \notin \Gamma_{n,k} } } }
\nonumber \\
&\leq& \prod_{i=1}^r \bE{ \left ( \left ( 1 - p \right )^{- \1{ k
\notin S_{i} } } \right )^{\1{ i \notin \Gamma_{n,k} } } }
\label{eq:NAwith(1-p)Inverted} \\
&=& \prod_{i=1}^r \bE{ \left ( 1 - p \right )^{-\1{ i \notin
\Gamma_{n,k} } } }^{\1{ k \notin S_{i} }} \nonumber
\end{eqnarray}
where (\ref{eq:NAwith(1-p)Inverted}) follows from the negative
association of the rvs (\ref{eq:NegativeAssociation1}) -- Use
(\ref{eq:NegativeAssociationConsequence}) and note that
\[
\left ( 1 - p \right )^{- \1{ k \notin S_{i} } } \geq 1, \quad i
=1, \ldots , r .
\]

Next we observe that for each $i=1, \ldots , r$, we have
\begin{eqnarray}
\lefteqn{ \bE{ \left ( 1 - p \right )^{-\1{ i \notin \Gamma_{n,k}
} } } } & &
\nonumber \\
&=& \left ( 1 - p \right )^{-1} \bP{ i \notin \Gamma_{n,k} } +
\bP{ i \in \Gamma_{n,k} }
\nonumber \\
&=& \left ( 1 - p \right )^{-1} \left ( 1 - \frac{K}{n-1} \right )
+ \frac{K}{n-1}
\nonumber \\
&=& \frac{u_n(\theta)}{1-p}
\nonumber
\end{eqnarray}
whence
\[
\prod_{i=1}^r \bE{ \left ( 1 - p \right )^{-\1{ i \notin
\Gamma_{n,k} } } }^{\1{ k \notin S_{i} }} = \left (
\frac{u_n(\theta)}{1-p} \right )^{ \sum_{i=1}^r \1{ k \notin S_{i}
} } .
\]
Combining these observations readily yields
\begin{eqnarray}
\lefteqn{ G_{n,r} (S_1, \ldots , S_r; \theta) } & &
\nonumber \\
&\leq& \prod_{k=r+1}^n \left ( \frac{u_n(\theta)}{1-p} \right
)^{\sum_{i=1}^r \1{ k \notin S_{i} } }
\nonumber \\
&=& \left ( \frac{u_n(\theta)}{1-p} \right ) ^{\sum_{i=1}^r
\sum_{k=r+1}^n \1{ k \notin S_i } } .
\nonumber 
\end{eqnarray}
We finally obtain
\begin{eqnarray}
\lefteqn{ \bP{ B_{n,r}(\theta) \Big | \Gamma_{n,1}, \ldots ,
\Gamma_{n,r} } } & &
\nonumber \\
&\leq& (1-p)^{r(n-r)} \left ( \frac{u_n(\theta)}{1-p} \right )
^{\sum_{i=1}^r \sum_{k=r+1}^n \1{ k \notin \Gamma_{n,i} } }
\nonumber
\end{eqnarray}
and the desired conclusion
(\ref{eq:BoundOnConditionalProbabilityB}) follows. \myendpf

\section*{Acknowledgment}
This work was supported by NSF Grant CCF-07290.

\bibliographystyle{IEEE}


\end{document}